# Technical Report #KU-EC-12-2:

# Exploratory and Inferential Methods for Spatio-Temporal Analysis of Residential Fire Clustering in Urban Areas

Elvan Ceyhan[a], Kıvanç Ertuğay[b], Şebnem Düzgün[c]
[a] Department of Mathematics, Koç University, Istanbul, Turkey
Phone: +90 (212) 338-1845 Fax: +90 (212) 338-1559
E-mail: elceyhan@ku.edu.tr
[b]Department of Urban and Regional Planning, Selcuk University, Konya, Turkey
Phone: +90 (332) 223-2203
E-mail: kivancertugay@gmail.com
[c]Geodetic and Geographic Information Technologies, Middle East Technical University, Ankara, Turkey
Phone: +90 (312) 210-2668 Fax: +90 (312) 210-5822
E-mail: duzgun@metu.edu.tr

**Abstract**

The spatio-temporal analysis of residential fires could allow decision makers to plan effective resource allocations in fire management according to fire clustering levels in space and time. In this study, we provide guidelines for the use of various methods in detecting the differences in clustering patterns of fire and non-fire (i.e., background residential) locations and how these patterns change over time. As a preliminary analysis step, various exploratory data analysis methods, such as, intensity plots (i.e., kernel density estimates) are used. Moreover, the use of Diggle's $D$-function (a second order analysis technique) is proposed for detecting the clustering of residential fire locations (if any) and whether there is additional clustering (or regularity) in the locations of the fires compared to background residential pattern. A test for trend over time (in years, months, and weeks) of the fire location patterns are provided with a space-time interaction analysis by spatio-temporal $K$-function. Residential fire data from Çankaya Municipality of Ankara, Turkey is used as an illustrative example. The presented methodology is also applicable to residential fire data from similar urban settings.

## 1. Introduction

Each year, residential fires in urban areas incur various social and economic losses with hundreds of deaths, thousands of injuries and million dollars of property loss. Although it is not possible to avoid fires entirely, it is always possible to reduce their harms by effective fire management.

Fire management is a collection of activities that involves systematic analyses, planning, decision making, assignment and coordination of available resources to manage fire related risks and includes interrelated sub-phases such as prevention, preparedness, response and recovery. In the fire management literature, mainly there are four types of research areas. The first type is related to the location/allocation, real time dispatching and accessibility problems [see, e.g., 1,2 and 3]. The second type concentrates on developing fire related databases and/or decision support systems [4 and 5]. The third type deals with understanding cause and effect relationships between fire and several other covariates [6, 7 and 8]. Finally, the fourth type concentrates on modeling, experimentation, and simulation of fire occurrence, spread, loss and so on. [9, 10 and 11].

Recently, spatial, temporal, and spatio-temporal methods for analysis of fire clustering have become a focus of considerable attention. For example,. Pew and Larsen [12] examine the spatial and temporal pattern of human-caused wildfires (HCWs) in the temperate rainforest of Vancouver Island (VI). A Geographic Information System (GIS) is used to locate HCWs that occurred from 1950 through 1992, by dividing VI into 1 km × 1 km grid cells, and determining the climate and distances to various human-built infrastructures for each grid cell. They employ logistic regression to build a model that predicts the probability of HCW occurrence using spatial data regarding climate and the distance to human-built infrastructures and determine whether temporal variations in the number of fires, and the area affected by them, have been constant over time or not in different HCW probability classes. In another example, Podur et al. [13] analyze spatial distribution of forest fires caused by lightning between 1976 and 1998 in Ontario region by using spatial data analysis techniques such as kernel density estimation, nearest neighbor and the $K$-function to detect whether the patterns are clustered or random within the study area. More recently, Corcoran et al. [14] apply spatial analysis methods to explore spatial dynamics and patterns of fire incidents in an area of South Wales. In a further research, Corcoran et al. [15] apply spatio-temporal methods to understand the interaction between four principal fire incident categories, namely property, vehicle, secondary fires, and malicious false alarms. They employ simple line and circular plots for different periods of time (i.e., hourly, daily, and monthly) to investigate temporal patterns; cumulative sum technique coupled with the kernel density method to investigate spatial patterns; the co-map technique which illustrates the entire time period under study in a single visualization for the interaction of space and time. Corcoran et al. [15] indicate that application of the spatio-temporal analysis techniques for fire incidents has potential to inform policy makers from both a reactive, resource-allocation perspective and a more proactive perspective, such as spatial targeting of preventive measures. Asgary et al. [16] also applied temporal, spatial, and spatio-temporal analysis techniques and illustrate how the patterns of structural fire incidents in Toronto, Ontario, Canada from 2000 to 2006 vary with the time of the day, the day of the week, and the month of the year. For temporal analysis, they present the data in the form of hour, day, month and year, while for the spatial analysis, they use quadrant count method, kernel density estimation and nearest neighbor distance methods, and in spatio-temporal analysis, they use map animation, isosurface and co-map methods. They demonstrate that there are significant differences with respect to fire incidents over time. Moreover, they point out that the application of these methods can help decision makers take preventive measures over time and space and improve resource allocations in case of emergency.

Although there is extensive research in fire management literature, the research on integration of spatial, temporal, and spatio-temporal data analysis techniques into fire clustering are quite limited and have started to gain attention recently. A better understanding of spatio-temporal patterns could provide vital information for effective fire management such as planning of fire prevention and response actions in terms of risk identification, resource targeting and routing of fire personnel and equipment, allocation of preventive measures, and policy evaluation with strategies for reducing fire related deaths, injuries and property losses [14,16].

In this paper, in addition to some exploratory methods available in fire safety literature, we propose the use of Diggle's $D$-function [17] in the spatial analysis of the residential fires, provide a test for trend of fire patterns over time, and spatio-temporal $K$- function [18] for space-time interaction analysis. Diggle's $D$-function allows one to determine how much of the clustering of residential fires (if any) is due to the clustering of the residences, and whether there is additional clustering (or regularity) in the locations of the fires. In residential areas, if probability of occurrence of fire is constant for each residence, fire-clustering pattern would follow the same pattern as the residential clustering. In other words, in residential areas with more dwelling intensity, fire cases would also be more likely to occur. Therefore, it is very important to determine if the fire clustering pattern is different (more intense or more regular) than the underlying (background) residential pattern. Kernel estimation, nearest neighbor distance analysis and the $K$-function methods have the shortcoming of indicating fire clusters without sufficient consideration of background population, which are the residential areas in this case. Although population weighted kernel density estimation can be a solution to this problem, it is only an exploratory tool and does not provide the statistical significance of any decision. The proposed spatio-temporal analysis methods are illustrated using residential fire data from Çankaya, an urban province of Ankara, Turkey. The residential fire data was collected on years 1998, and 2005-2009 where in the last three years the month, week, and day of fire (starting from the beginning of the year) are also available. In the same region, locations of the residences which did not experience fire (referred to as non-fire cases, henceforth) are randomly sampled. The differences in patterns of fire and non-fire locations are identified and how these patterns change over time are investigated. The relative patterns of fire and non-fire cases are compared using Diggle's $D$-function, which indicates the additional regularity or clustering of fire cases compared to the non-fire cases. Moreover, intensity plots (i.e., kernel density estimates) of the fire locations and non-fire locations for data on all years (referred to as overall data, henceforth) and for each year are also performed. Test for trend over time in years, months, and weeks of the fire patterns are carried out and a space-time interaction analysis is performed by using spatio-temporal $K$- function of Rowlingson and Diggle [18], which provides better identification of spatio-temporal clusters than the conventional $K$- function.

As a final remark, we would like to emphasize that the focus of the manuscript is on the innovativeness. The primary aim of the study is not to evaluate a specific fire clustering pattern in a detailed manner but to provide guidelines to the decision makers for the usage of various spatio-temporal data analyses techniques in understanding fire clustering patterns.

## 2. Analysis of Spatial and Spatio-Temporal Patterns

In general, a point pattern like the spatial distribution of fire locations may exhibit one of the three types of patterns, namely, complete spatial randomness (CSR), clustering, or regularity or a mixture of them in an appropriately defined region [19]. Hence determining which type of pattern exists in any given fire location distribution requires analyzing the spatial distribution of points in global (first-order properties) and local (second-order properties) scale. The following subsections provide the background on methods of spatial and spatio-temporal point pattern analysis.

### 2.1. Spatial Clustering Methods

#### 2.1.1. Local Intensity and Intensity Ratios

Diggle [20] and Berman and Diggle [21] estimate a spatially smooth local intensity of the points, $\hat{\lambda}(x)$, by a kernel method which employs the quadratic kernel function:

$$f(x) = \begin{cases} (1-u^2/2)^2 & -\sqrt{2} \le u \le \sqrt{2}, \\ 0 & \text{otherwise} \end{cases} \tag{1}$$

and the resulting estimate is $\hat{\lambda}(x) = h_0^{-1} \sum_{i=1}^{n} f(d_i / h_0)$ (2)

where $d_i$ is the distance from point *i* to *x*.

A first description of a spatial pattern can be performed by the spatial statistical density, which is proportional to the spatial intensity. The spatial density and intensity are parts of the first-order properties, because they measure the distribution of events (here locations of fire and non-fire cases) in the study region. On the other hand, the second-order properties measure the strength and types of interaction between events of the point process(es). Therefore, the second-order properties are particularly interesting if one wants to study the clustering or interaction between events. Informally, the second-order intensity of two events i and j reflects the probabilities of any pair of events occurring in the vicinities of i and j, respectively. The most common method used for second-order analysis is Ripley's $K$-function, which is defined as

$$K_{ii}(s) = \lambda^{-1} E[\text{\# of extra events within distance s of a randomly chosen event}] \tag{3}$$

with λ being the density (number of fires per unit area) of events and $K$-function is estimated as

$$\hat{K}_{ii}(s) = \hat{\lambda}^{-1} \sum_{i} \sum_{i \neq j} w(i, d_{ij}) I(d_{ij} < s) / N \tag{4}$$

where $\hat{\lambda} = N / A$ is an estimate of density ($N$ is the observed number of points and $A$ is the area of the study region), $d_{ij}$ is the distance between points $i$ and $j$, $I(\cdot)$ is the indicator function, $w(i, d_{ij})$ is the proportion of the circumference of the circle centered at point $i$ with radius $d_{ij}$ that falls in the study area, which corrects for the boundary effects (See [17] for more detail). When the null case is the random labeling (RL) of points from an inhomogeneous Poisson process (like the case of this paper), Ripley's $K$- or $L$-functions in the general form are not appropriate to test for the spatial clustering of the cases [22]. However, Diggle [17] suggests a version based on Ripley's univariate $K$-function as

$$D(s) = K_{11}(s) - K_{22}(s) \tag{5}$$

where $K_{11}(s)$ is usually Ripley's $K$- function for the class of interest and $K_{22}(s)$ is Ripley's $K$-function for the background class. In our setup, $K_{22}(s)$ measures the degree of spatial clustering of non-fires (i.e., the background population at risk), while $K_{11}(s)$ measures this same spatial clustering plus any additional clustering due to fire patterns. In this formulation, "no spatial clustering" is equivalent to RL of fires and non-fires on the locations in the sample, which implies $D(s) = 0$. The test statistic $D(s)$ is estimated by

$$\hat{D}(s) = \hat{K}_{11}(s) - \hat{K}_{22}(s)\,. \tag{6}$$

## 2.2. Spatio-Temporal Clustering Methods

### 2.2.1. Second-Order Analysis of Spatio-Temporal Clustering of Fire Cases

The theory of point processes on general spaces is well established [e.g., 23]. For the purpose of analyzing the distribution of fires, the points of the process, termed events (i.e., fire cases), are a countable set of points $(x_i, t_i)$ where $x_i \in R^2$ represents the spatial location of the $i^{th}$ event and $t_i \in R$ its associated reference time. For a stationary process, the intensity $\lambda$ is identified as the expected number of events per unit space per unit time, and the reduced second moment measure or spatio-temporal $K$-function, as

$$K(s,t) = \lambda^{-1} E\left[\#\text{ of extra events within distance s and time t of an arbitrary event}\right]. \tag{7}$$

For a spatio-temporal homogeneous Poisson process, in which the spatial and temporal components are independent homogeneous Poisson processes on $R^2$ and $R$, respectively,

$$K(s,t) = 2\pi s^2 t. \tag{8}$$

Equation (8) can be used as a benchmark against which the second order properties of a spatio-temporal process. If these component processes are independent, it then follows that the $K$-function in (8) factorizes as

$$K(s,t) = K_1(s)K_2(t) \tag{9}$$

In Equation (9), $K_1(\cdot)$ and $K_2(\cdot)$ are the $K$-functions of the spatial and temporal component processes,

$$K_1(s) = \lambda_1^{-1} E[\text{\# of extra events within distance s an arbitrary event}] \tag{10}$$

and

$$K_2(t) = \lambda_2^{-1} E[\text{\# of extra events within time t an arbitrary event}], \tag{11}$$

respectively.

In this study, the spatio-temporal interaction in the distribution of the residential fire cases is analyzed by using the function of spatial and temporal separation. The main question of interest is whether fire cases, which are close in space, are also close in time or not. If they are close in time, then it can be deduced that the fire distribution exhibits spatio-temporal clustering.

In most cases fire data might show both spatial clustering, reflecting a non-uniform geographical distribution of the population at risk over the region in question, and temporal clustering, For example the fire occurrences might occur more frequently in a particular time period. Unless there is a series of spatially and temporally localized increases in the fire incidences, it can be expected that these spatial and temporal effects operate independently. For this reason, the detection of spatio-temporal clustering constitutes a preliminary step for a more detailed analysis of fire in question. In this paper, the levels of spatio-temporal clustering of fire cases are estimated as a function of spatial and temporal separation. The analysis is based on the second-order properties of a spatio-temporal point process of fire cases.

### 2.2.2. Estimation of Second-Order Properties

Let $\{(x_i, t_i) : i = 1,...,n\}$ denote the locations and times of all events within a spatio-temporal region $A \times (0,T)$. Let $d_{ij} = \|x_i - x_j\|$ and $u_{ij} = |t_i - t_j|$ be the spatial and temporal separations of the $i^{th}$ and $j^{th}$ events. Let $w_{ij}$ be the reciprocal of the proportion of the circumference of the circle with centre $x_i$ and radius $d_{ij}$ which lies within $A$. Let $v_{ij} = 1$ if both ends of the interval of length $2u_{ij}$ and centre $t_i$ lie within $(0,T)$, $v_{ij} = 2$ otherwise. Then, using $I(\cdot)$ to denote the indicator function, approximately unbiased estimators for $K(s,t)$, $K_1(s)$ and $K_2(t)$ are

$$\hat{K}(s,t) = |A| T \{n(n-1)\}^{-1} \sum_{j \neq i} w_{ij} v_{ij} I(d_{ij} < s) I(u_{ij} < t), \tag{12}$$

$$\hat{K}_1(s) = |A| \{n(n-1)\}^{-1} \sum_{j \neq i} w_{ij} I(d_{ij} < s), \tag{13}$$

and

$$\hat{K}_2(t) = T\{n(n-1)\}^{-1}\sum_{j\neq i} v_{ij} I(u_{ij} < t). \tag{14}$$

The three equations above are the obvious extensions to spatial-temporal processes of Ripley's estimator [24,25] for a purely spatial point process. The weights $w_{ij}$ constitute an important correction for edge effects in estimating the spatial second order properties. The corresponding weights $v_{ij}$ play the same role for the temporal properties, where edge effects are much less important in practice. See, for example [26].

### 2.2.3. Diagnostics for Spatio-temporal Clustering

Estimated $K$-functions are widely accepted as useful tools for analyzing spatial point patterns [19, 17]. In this paper, simple diagnostic procedures for analyzing possible dependence between the spatial and temporal components of the underlying spatial-temporal point process are adopted.

First, the following functions are considered:

$$\hat{D}(s,t) = \hat{K}(s,t) - \hat{K}_1(s)\hat{K}_2(t) \tag{15}$$

and

$$\hat{D}_0(s,t) = \hat{D}(s,t)/[\hat{K}_1(s)\hat{K}_2(t)]. \tag{16}$$

Equation (15) is based on Equation (9) and yields another benchmark relative to which second order dependence between the spatial and temporal component processes can be assessed. A perspective (i.e., three-dimensional) plot of the surface $\hat{D}(s,t)$ gives information on the scale and nature of the dependence between the spatial and temporal components, and constitutes our first suggested diagnostic for space-time clustering. Note that $\hat{D}(0,0) = 0$. Also, as it will be shown in further sections, the sampling fluctuations in $\hat{D}(s,t)$ typically increase with s or t and the useful information in $\hat{D}(s,t)$ is therefore confined to values of $s$ and $t$ which are small relative to the spatial and temporal dimensions of $A\times(0,T)$.

In general, the sampling distributions of the quantities $\hat{D}(s,t)$ are intractable, but progress is possible if the spatial and temporal component processes are independent and conditionally on the realizations of each component. In this null case the sampling distribution of $\hat{D}(s,t)$ is the distribution induced by random permutation of the times $\{t_i : i = 1,...,n\}$ for a fixed set of locations $\{x_i : i = 1,...,n\}$, or vice versa [27].

Let $V(s,t)$ denote the variance of $\hat{D}(s,t)$ from [27], and write $\hat{K}_0(s,t) = \hat{K}_1(s)\hat{K}_2(t)$. Then, a third diagnostic for space-time interaction is a plot of the standardized quantities $R(s,t) = \hat{D}(s,t) / \sqrt{V(s,t)}$ against $\hat{K}_0(s,t)$, analogous to a plot of standardized residuals against fitted values in regression modeling. The main advantage of this diagnostic plots over the plot of $\hat{D}(s,t)$ or $\hat{D}_0(s,t)$ is that it is two-dimensional rather than three dimensional and easier to visualize. A corresponding disadvantage is that the spatial and temporal scales are no longer explicit.

#### 2.2.4. Tests for Space-Time Interaction

A discrete approximation to the integral of the standardized residual surface given below can be used,

$$U = \sum_s \sum_t R(s,t) \tag{17}$$

Significantly, positive or negative values indicate positive or negative space-time interaction, respectively. Whatever statistic $U$ is chosen, it makes sense to confine attention to values of $s$ and $t$ which are small relative to the dimensions of the region on which the data are observed. For an approximate test of significance, based on the above test statistic, the null variance of $U$ from [27] can be evaluated, and it refers to the standardized statistic $U / \sqrt{Var(U)}$ to critical values of $N(0,1)$. An exact alternative, applicable to statistic $U$, is a Monte Carlo test, in which the observed value $u_1$ of $U$ is ranked amongst values $u_2,...,u_m$ generated by re-computing the $\sum_s \sum_t R(s,t)$ after each of $m-1$ independent random permutations of the reference times; if $u_1$ ranks $k^{th}$ largest (or smallest) the one-sided attained significance level is $k/m$. Both the Monte Carlo test and the approximate normal (i.e., Gaussian) test based on Equation (17) are computation-intensive, but the latter can be simplified by computing $V(s,t)$ only for a coarse grid of values of $s$ and $t$, and interpolating.

## 3. Case Study Implementation

### 3.1. Data Description

The city of Ankara is the second most populated city and the capital of Turkey with a population of more than 4 million people. Ankara is governed by a “greater metropolitan municipality”, including eight main metropolitan districts, which are Çankaya (the case study area), Yenimahalle, Keçiören, Altındağ, Mamak, Sincan, Etimesgut and Gölbaşı (see Figure 1).

Çankaya district is one of the oldest districts of Ankara with an area of 12,700 hectares and approximate population of 800,000 people according to the results of 2007 census [28].

In Turkey, fire protection and prevention, taking necessary measures against fires, rescuing the citizens in accidents and other emergency situations and organizing training programs against fire threat are assigned to municipalities by Law of Municipalities no 1580. In metropolitan cities like Ankara, this responsibility is executed by Metropolitan Municipalities pursuant to Law of Metropolitan Municipalities no 3030.

Fire Department of Ankara Metropolitan Municipality is comprised of the following units: head office of Ankara Municipality Fire Department in İskitler (Ankara Büyükşehir Belediyesi İtfaiye Daire Başkanlığı) is the most authorized unit of Ankara fire department and reports to the Mayor of Metropolitan Municipality of Ankara. There are 14 fire brigades reporting to Ankara Metropolitan Municipality Head office of Ankara Fire Department: Sincan, Batıkent, Keçiören, Altınpark, Siteler, Kurtuluş, Kale, Kayaş, Esat, Köşk, Gölbaşı, AŞTİ, and Esenboğa. Four of the 14 fire brigade units, namely Kale, Köşk, AŞTİ and Esenboğa fire departments are special departments and only responsible from their specific areas which are historical Kale region, Presidential Palace, Ankara Intercity Bus Terminal, and Esenboğa airport, respectively (Figure 2a). The other 10 fire brigades are responsible from the fires that take place in their fire responsibility zones of Sincan, Batıkent, Keçiören, Altınpark, Siteler, Kurtuluş, Kayaş, Esat, Gölbaşı, and İskitler (head office) as shown in Figure 2b.

In particular, Çankaya district is under the responsibility of six different fire brigade services which are İskitler (head office), Kurtuluş, Esat, Gölbaşı, AŞTİ, and Köşk fire brigades (Figure 2c). İskitler, Kurtuluş, Esat and Gölbaşı fire brigades are the main units that serve Çankaya district and the other two brigades of AŞTİ and Köşk have special responsibilities as described above. The fire responsibility zones do not follow the district borders since they are constructed based on transportation network (i.e., the road and street network) in Ankara.

The residential fire data of Çankaya district is obtained from Head Office of Ankara Fire Department in tabular format by special permission from Ankara Metropolitan Municipality. The obtained data consist of the address information of all residential fires reported to Head Office of Ankara Fire Department that occur inside the administrative borders of Çankaya district in 1998 and 2005-2009. The locations of fire cases are residences, dwellings or shops. For the time component, the reference time is the date of fire and for each of years 2007-2009 (inclusive) the month, week, and day of the fire starting from the beginning of the year are available. The fire data obtained in tabular format are geo-coded and converted to point features in GIS environment. In geo-coding phase, it was only possible to geo-code nearly 52% of the 2007-2009 data, because of the address matching limitations of the GIS software. However this problem does not affect the results considerably as the primary aim of the study is not to perform an extensive case study of a real life data, but provide a methodology to understand the residential fire clustering pattern relative to the non-fire residential pattern.

### 3.2. Spatial Clustering Analysis of Residential Fire Cases

The first step in such analysis is to perform exploratory data analysis to visually inspect the point pattern. In Figures 3 and 4, the scatter plots of the locations of overall non-fire cases and yearly residential fire cases are presented. Notice that the fire and non-fire cases seem to exhibit similar but not identical clustering patterns. The non-fire cases represent the distribution of the residential units, which seems to be non-uniform in the region of interest as expected. In fact, due to physical restrictions such as lakes, rivers, rocky or uneven terrains, and social restrictions or public regulations, usually locations of residences are different from the CSR pattern in the re-gion. See, for example, Figure 5 for Ripley's (modified) $L$-function for non-fire locations, which indicate significant clustering for distances from 0-500 meters. Moreover, one might also expect that the higher the residential intensity (number of houses per unit area), the higher the risk of residential fire, and hence the higher the fire case intensity (number of fires per unit area). Therefore, a clustering analysis of fire cases only may be misleading about the dynamics and the risk of fire distribution pattern. For example, if every residential unit has the same risk of fire, fire clustering and residential clustering will have the same pattern. Hence it is more reasonable to analyze the clustering of fire cases with respect to the residential clustering present (in the background). In case of excess clustering of fire cases (i.e., if fire cases are more clustered compared to non-fire residences), then we can conclude that the houses in that neighborhood have higher risk of fire (compared to the rest of the region). On the other hand, if fire cases are less clustered than the non-fire residences, then houses in that neighborhood have lower risk of fire. This inference might have important implications for precautionary measures against fire or rates of fire insurance.

In Figure 3, the locations of the non-fire cases are plotted for all years combined and in Figure 4 fire cases for each year are plotted. The plots for each year apparently shows a different trend in fire (at least for some years) and there seems to be some differences between the clustering of fire cases compared to non-fire cases. In years 1998, 2005-2007, fire cases tend to occur mostly on dense residential areas. However, in years 2008 and 2009, fire clustering also occurs in southwest part of the region where the residential intensity is lower. A more rigorous formal analysis is presented in Sections 3.3 and 3.4.

With the kernel estimation in Equation (2), the intensity estimates of fire and non-fire cases are obtained and presented in Figure 6. Ignoring the time effect, it is observed that the overall fire and non-fire intensities follow a similar but not the same trend. There seems to be higher intensity (or clustering) in fire and non-fire clustering in north-eastern part of the region. In Figure 6 the fire intensity plots for each year are also presented. By visual inspection, in 2005 and 2006 fire and non-fire intensities are similar; while in the other years they seem to be different. Observe that the fire clustering pattern more or less follow the trend of residence clustering, but perhaps with different intensities. Furthermore, in 1998, there seems to be one major fire cluster, but in other years there seems to be multiple (possibly intersecting) major clusters.

### 3.3. Second-Order Analysis of Residential Fire Cases

In Figure 7, we present Ripley's $K$-function for overall fire and non-fire patterns and Diggle's $D$-function for the overall data (ignoring the time effect). Notice that level of the residential clustering is significantly higher than the benchmark CSR pattern for all the inter-point distances considered (i.e., 0-100 m), and the same holds for fire clustering also. However, the level of fire clustering, although follows the same trend, is significantly higher than non-fire residential clustering which is also seen in the plot of Diggle's $D$-function, since $\hat{D}(s)$ curve is above the upper confidence limit. That is, the fire pattern exhibit a higher level of clustering compared to the non-fire cases. In Figure 8, Diggle's $D$-function for each year is presented. Observe that for 1998, 2005, and 2007-2009, the fire locations seem to be (significantly) more clustered at all the distances considered (namely, from $s = 0$ to 100 m). On the other hand, in 2006, the clustering of fires does not significantly differ from the clustering of residences (i.e., non-fire cases) at small scales (for $s$ up to around 80 m), and for larger distances, fire locations seem to be more clustered than the residences. This is more or less in agreement with the visual inspection of intensity plots in Figure 6. Furthermore, these results suggest that fire clustering pattern changes over time (hence suggest a temporal trend or clustering as well) which is analyzed in Section 3.4.

### 3.4. Analysis of Temporal Clustering of Fire Cases

In the analysis of the temporal clustering of residential fire cases, first, the frequency histogram of fire cases by year in Figure 9 (left) are plotted, where it is observed that the frequency of the fire cases tends to reduce by year with largest being about 450 cases in 1998. The fire incidence frequencies by month for the years 2007-2009 combined in Figure 9 (right) are also considered. Notice that fire cases are most frequent in February, April, July, and December, whereas least frequent in September. The fire incidence frequencies by month for each year between 2007 and 2009 (inclusive) are plotted in Figure 10. Fire cases are most frequent in February and December in 2007; January, February, July and December in 2008, and March, June, and December in 2009. In these years, fire cases are less frequent in September. Therefore, one might conclude that fire occurs more frequently in colder months when people use coal or natural gas stoves (unless central heating) or in hotter months following perhaps a period of draught.

### 3.5. Spatio-Temporal Analysis of Residential Fire Cases

In this section, the space-time clustering analysis of [18] for the residential fire data is performed. In Figure 5, the locations of fire and non-fire cases (pooled over all the years) are given. Observe that the clustering of fire cases seem to differ from year to year.

#### 3.5.1. Spatio-Temporal Analysis for Time Measured in Years

In spatio-temporal analysis, the $K$-function in space, time, and space-time is estimated. The spatial $K$-function for distances from 0-100 m and temporal $K$-function for time differences in years from 0-11 years are plotted in Figure 11. Observe that the fire locations tend to be clustered in time and space. Furthermore, the values of $\hat{K}(s,t)$ tend to increase as time and distance increases, and the substantial jump for the recent years suggests that space-time interaction occurs in the fire distribution in this county. The standard error for each pair of $(s,t)$ values is stored in a matrix and plotted as a color-coded grid plot in Figure 12. Notice that the largest estimates of the standard error occurs for distances around 90-100 m and time differences around 5 years. That is, the estimated $\hat{K}(s,t)$ has more variability at 5-year time differences and for distances between 90-100 m. In the spatial component, this is no coincidence, because as the inter-point distance s increases the standard error of $\hat{K}(s)$ increases. On the other hand, such a trend does not occur in time differences, standard error for the spatio-temporal $K$-function $\hat{K}(s,t)$ tends to be larger for large inter-point distances considered and moderate time differences.

The diagnostic plots for spatio-temporal clustering are presented in Figure 13. The top left is the locations of the fire cases (all years combined). The top right presents the perspective plot of $\hat{D}(s,t)$. Notice that $\hat{D}(s,t)$ values increase in distance s, but much faster for smaller time scales. A scatter plot of $R(s,t)$ versus $\hat{K}_0(s,t) = \hat{K}(s)\hat{K}(t)$ is presented in Figure 13 (see the residual plot in bottom left). This residual plot strongly suggests the presence of space-time interaction in the data for smaller $\hat{K}_0(s,t)$ values, since the standardized residuals $R(s,t)$ are almost all positive with an average close to 6, whereas in the absence of space-time interaction $R(s,t)$ would have expected value 0 and variance 1. The larger $R(s,t)$ values at smaller $\hat{K}_0(s,t)$ values suggest that the strongest interaction is at the smaller spatial and/or temporal scales. That is, the level spatial clustering changes in time as well and largest spatio-temporal clustering occurs at large spatial scales and small time scales (see plot of $\hat{D}(s,t)$ in top right). Furthermore, the residual plot is suggestive of nine groupings by year. A Monte-Carlo test is also performed for space-time interaction where the sum of residuals as a test statistic is used, randomly permuting the times of the set of points and recomputing the test statistic for a number of simulations (See [27] for details). The observed value of the test statistic was $u_1 = 253199791.0$, while the values $u_2,\ldots,u_{1000}$ from 999 permutations of the times ranged from -59396644 to 72773452, so the test statistic of the case seems highly significant ( $p = 0.001$) which implies significant space-time interaction for the time at year level. See Figure 13 (bottom right). This implies that the spatial clustering pattern changes over year.

#### 3.5.2. Spatio-Temporal Clustering Analysis with Time Measured in Days, Weeks, and Months for Year 2007

The spatial and temporal $K$-functions in year 2007 for the time differences measured in months, weeks, and days are also obtained and given in Figure 14 together with the 95% confidence bands. Notice that there seems to be no significant temporal clustering at the month, week, and day levels, except perhaps for time differences around 30 weeks or 210 days, where for these time differences temporal clustering is smaller compared to the one under temporal randomness.

The diagnostic plots for spatio-temporal clustering by month are presented in Figure 15. Notice that $\hat{D}(s,t)$ values increase in distance s, but much faster for 6-8 month time differences (see Equation (15) for definition of $\hat{D}(s,t)$). The residual plot in bottom left strongly suggests a presence of space-time interaction in the data for smaller $\hat{K}_0(s,t)$ values, since the standardized residuals $R(s,t)$ are almost all positive with an average close to 3. The larger $R(s,t)$ values at smaller $\hat{K}_0(s,t)$ values suggest that the strongest interaction is at the smaller spatial and/or temporal scales. The Monte-Carlo test for space-time interaction is performed. The observed value of the test statistic was $u_1 = 869801959$, while the values $u_2,\ldots,u_{1000}$ from 999 permutations of the times ranged from -1086691779 to 1066202930, hence the test statistic seems highly significant ( $p = 0.005$) which implies significant space-time interaction for the time at the month level. That is, spatial clustering pattern changes over month in 2007. Hence different design of fire management strategies may be appropriate for different months of the year and parts of the study region. See Figure 15 (bottom right). The color-coded grid plot of the standard error values for the $(s,t)$ values in Figure 16 suggests that the largest standard error estimates occur for distances 90-100 m, and within the middle part of the time differences (4-6 months or 12-25 weeks). That is, the estimated $\hat{K}(s,t)$ has more variability for distances between 90-100 m and for time differences about 5 months apart. The trend is similar in standard error estimates for 2008 and 2009, hence not presented.

The diagnostic plots for spatio-temporal clustering by week are presented in Figure 17. The trend in $\hat{D}(s,t)$ values and residual plot is similar to those in Figure 15. The Monte-Carlo test for space-time interaction yields an observed value of the test statistic as $u_1 = 2344377975$, while the values $u_2,\ldots,u_{1000}$ from 999 permutations of the times ranged from -17061211844 to 14744768337, so our test statistic seems almost significant ($p = 0.330$) which implies lack of space-time interaction for the time at the week level. Therefore, it can be inferred that spatial clustering pattern does not change significantly over week.

The diagnostic plots for spatio-temporal clustering by day are not presented. The Monte-Carlo test for space-time interaction yields an observed value of the test statistic as $u_1 = 216229193990$, while the values $u_2,\ldots,u_{1000}$ from 999 permutations of the times ranged from -800188507518 to 672789092223, so our test statistic does seem to be significant ($p = 0.163$) which implies lack of space-time interaction for the time at the day level. Hence, spatial clustering pattern does not change significantly over day.

### 3.5.3. Spatio-Temporal Clustering Analysis with Days, Weeks, and Months for Year 2008

The spatial and temporal $K$-functions in year 2008 for the time differences measured in month, week, and day is presented in Figure 18. Notice that there seems to be no significant temporal clustering at time differences measured in the month, week, and day levels, except possibly around 30 weeks or 210 days.

The diagnostic plots for spatio-temporal clustering by month are presented in Figure 19. Notice that $\hat{D}(s,t)$ values increase in distance s, but much faster for 4-8 month time differences. The residual plot in bottom left suggests a presence of space-time interaction in the data for smaller $\hat{K}_0(s,t)$ values, since the standardized residuals $R(s,t)$ are almost all positive with an average close to 1. The larger $R(s,t)$ values at smaller $\hat{K}_0(s,t)$ values suggest that the strongest interaction is at the smaller spatial and/or temporal scales. The Monte-Carlo test for space-time interaction is also performed. The observed value of the test statistic was $u_1 = 464413061$, while the values $u_2,\ldots,u_{1000}$ from 999 permutations of the times ranged from -793370610 to 1013279710, so our test statistic seems almost significant ($p = 0.080$) which implies almost significant space-time interaction for the time at the month level in 2008. See Figure 19 (bottom right).

The diagnostic plots for spatio-temporal clustering by week are presented in Figure 20. The trend in $\hat{D}(s,t)$ values and residual plot is similar to those in Figure 19. The Monte-Carlo test for space-time interaction yields an observed value of the test statistic as $u_1 = 3987436633$, while the values $u_2, \ldots, u_{1000}$ from 999 permutations of the times ranged from -15438047524 to 14821783616, so the test statistic does not seem to be significant ($p = 0.185$) which implies no significant space-time interaction for the time at the week level in 2009.

The diagnostic plots for spatio-temporal clustering by day are not presented. The Monte-Carlo test for space-time interaction yields an observed value of the test statistic as $u_1 = 197274290323$, while the values $u_2, \ldots, u_{1000}$ from 999 permutations of the times ranged from -732937733875 to 690071950257, so the test statistic is not significant ($p = 0.179$) which implies lack of space-time interaction for the time at the day level in 2009.

#### 3.5.4. Spatio-Temporal Clustering Analysis with Days, Weeks, and Months for Year 2009

The spatial and temporal $K$-functions in year 2009 are obtained for the time differences measured in month, week, and day in Figure 21. Notice that there seems to be no significant temporal clustering at the month, week, and day levels, except for around 30 weeks or 210 days.

The diagnostic plots for spatio-temporal clustering by month are presented in Figure 22. Notice that $\hat{D}(s,t)$ values increase in distance $s$, but much faster for 6 and 10 month time differences. The residual plot in bottom left suggests a lack of space-time interaction in the data, since the standardized residuals $R(s,t)$ are scattered around 0. The Monte-Carlo test for space-time interaction is performed. The observed value of the test statistic was $u_1 = 77109709$, while the values $u_2, \ldots, u_{1000}$ from 999 permutations of the times ranged from -861038600 to 882224074, so our test statistic seems insignificant ($p = 0.377$) which implies lack of significant space-time interaction for the time at the month level in 2009. See Figure 22 (bottom right).

The diagnostic plots for spatio-temporal clustering by week are presented in Figure 23. The trend in $\hat{D}(s,t)$ values and residual plot is similar to those in Figure 22. The Monte-Carlo test for space-time interaction yields an observed value of the test statistic as $u_1 = 3935490663$, while the values $u_2, \ldots, u_{1000}$ from 999 permutations of the times ranged from -16865949844 to 12322120151, so our test statistic seems insignificant ($p = 0.180$) which implies lack of significant space-time interaction for the time at the week level in 2009.

The diagnostic plots for spatio-temporal clustering by day are not presented. The Monte-Carlo test for space-time interaction yields an observed value of the test statistic as $u_1 = 194269350023$, while the values $u_2, \ldots, u_{1000}$ from 999 permutations of the times ranged from -800334162329 to 589557776077, so our test statistic seems mildly significant ($p = 0.177$) which implies lack of significant space-time interaction for the time at the day level in 2009.

## 4. Discussion and Conclusions

The main goal of this study is to provide some methodology for the spatio-temporal analysis of residential fire clustering patterns relative to the background (non-fire) residential pattern. Along this line, Çankaya district of Ankara is used as an illustrative example. This work could help decision makers to detect problematic regions in terms of spatio-temporal patterns and concentrate on these specific regions at specific times to develop policies and strategies for fire prevention and management. This research can also help decision makers to reduce the damages due to fires. Usually fire brigades can miss the spatio-temporal fire patterns due to their heavy workload. The decision makers can catch the important clues that are underlying the spatio-temporal distribution patterns of fire locations.

The use of spatio-temporal analysis techniques enables one to understand of the general trends about patterns of fire locations and thereby help the decision maker focus on a quarter or a street rather than the whole responsibility zone. That is, understanding the regions where there is evidence of clusters, which have previously not been detected, helps redesign the fire prevention or management efforts in specific regions rather than the whole urban area.

- Strengthening or building new fire stations closer to problematic areas,
- Distribution of current fire staff and equipment into smaller and widespread stations considering critical areas,
- Replacement or deployment of critical facilities if necessary,
- Improving the fire resistance of buildings and physical structures, fire communication opportunities in problematic areas,
- Legislative obligations for improving fire resistance of buildings by using fire resistant materials in constructions; building fire warning systems (alarm) and fire sprinkler systems (automatic fire extinguishment systems) in constructions (especially for problematic areas); fire insurance in critical areas.
- Overlaying critical areas with other additional urban related maps such as fire service accessibility maps, land use/cover maps, population density maps etc so as to extract additional fire related information."

In this article, analysis of spatio-temporal patterns of residential fires are provided at four stages: (i) exploratory analysis of fire clustering by scatter plots of fire versus non-fire locations and by intensity plots, (ii) spatial analysis of fire and non-fire patterns by Diggle's $D$-function, (iii) temporal analysis of fire clustering by temporal version of Ripley's $K$-function, and (iv) analysis of spatio-temporal interaction by spatio-temporal $K$-function and Monte Carlo hypothesis testing. However, it is important to stress that the evaluation of the exploratory analysis such as intensity plots are intuitive processes and depends on some user defined parameters. Therefore, in most of the cases the exploratory analysis may be insufficient and it may be required to go further to test various hypotheses or build models to explain the observed fire patterns. Hence second order analysis of fire and non-fire patterns by Diggle's $D$-function serves for this purpose and provides a Monte Carlo test for space-time interaction. The predictive modeling of spatio-temporal fire patterns is a topic of ongoing research. The results of this study are site specific; however, the methodology followed can be adapted to any urban location similar to ours.

## Acknowledgments

We would like to thank an anonymous associate editor and two referees, whose constructive comments and suggestions greatly improved the presentation and flow of the manuscript. E. Ceyhan was supported by the research agency TUBITAK via the 1001 Grant # 111T767.

# FIGURES

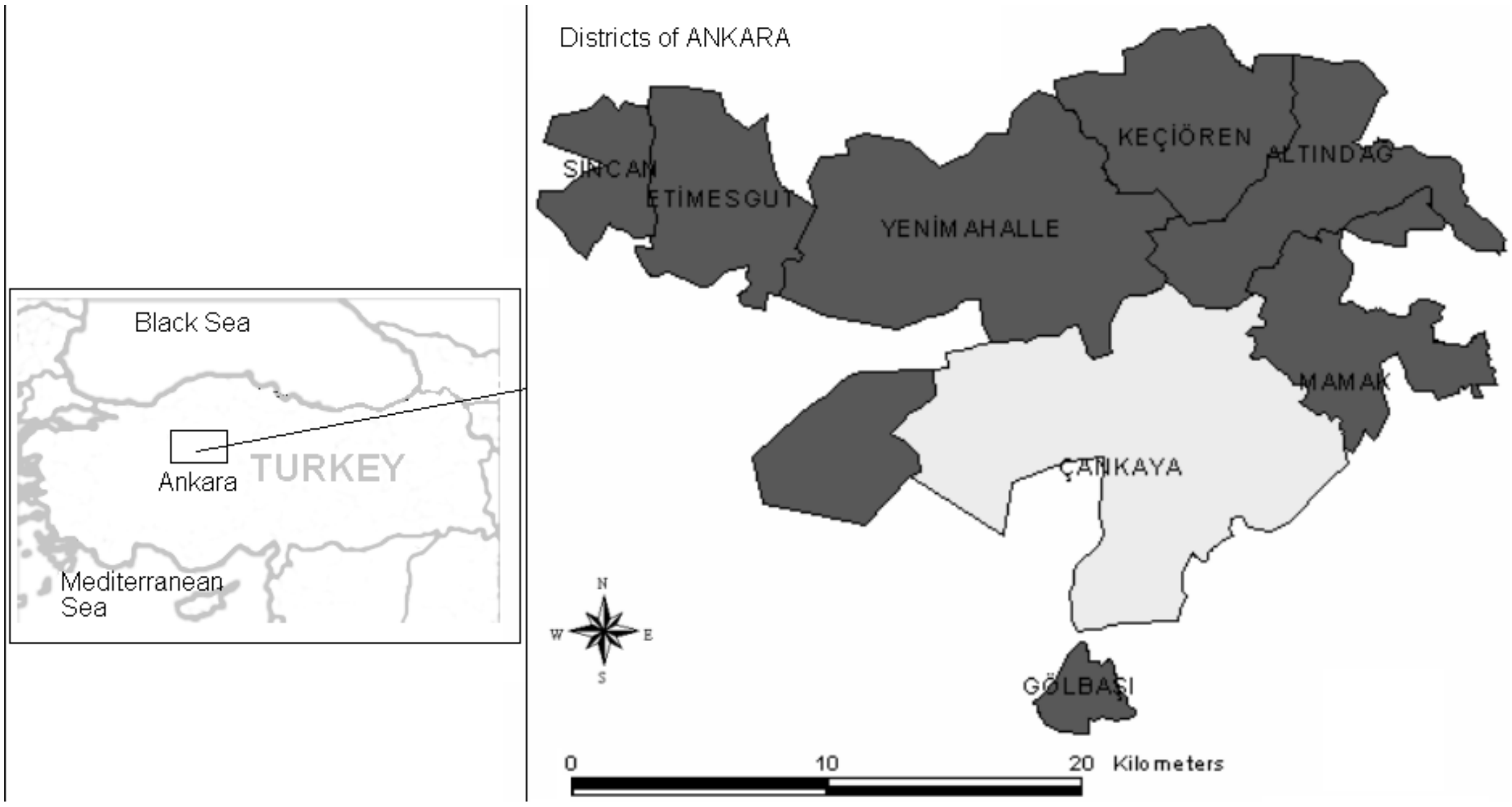

Figure 1: Study area (light gray region), Çankaya district, Ankara, Turkey

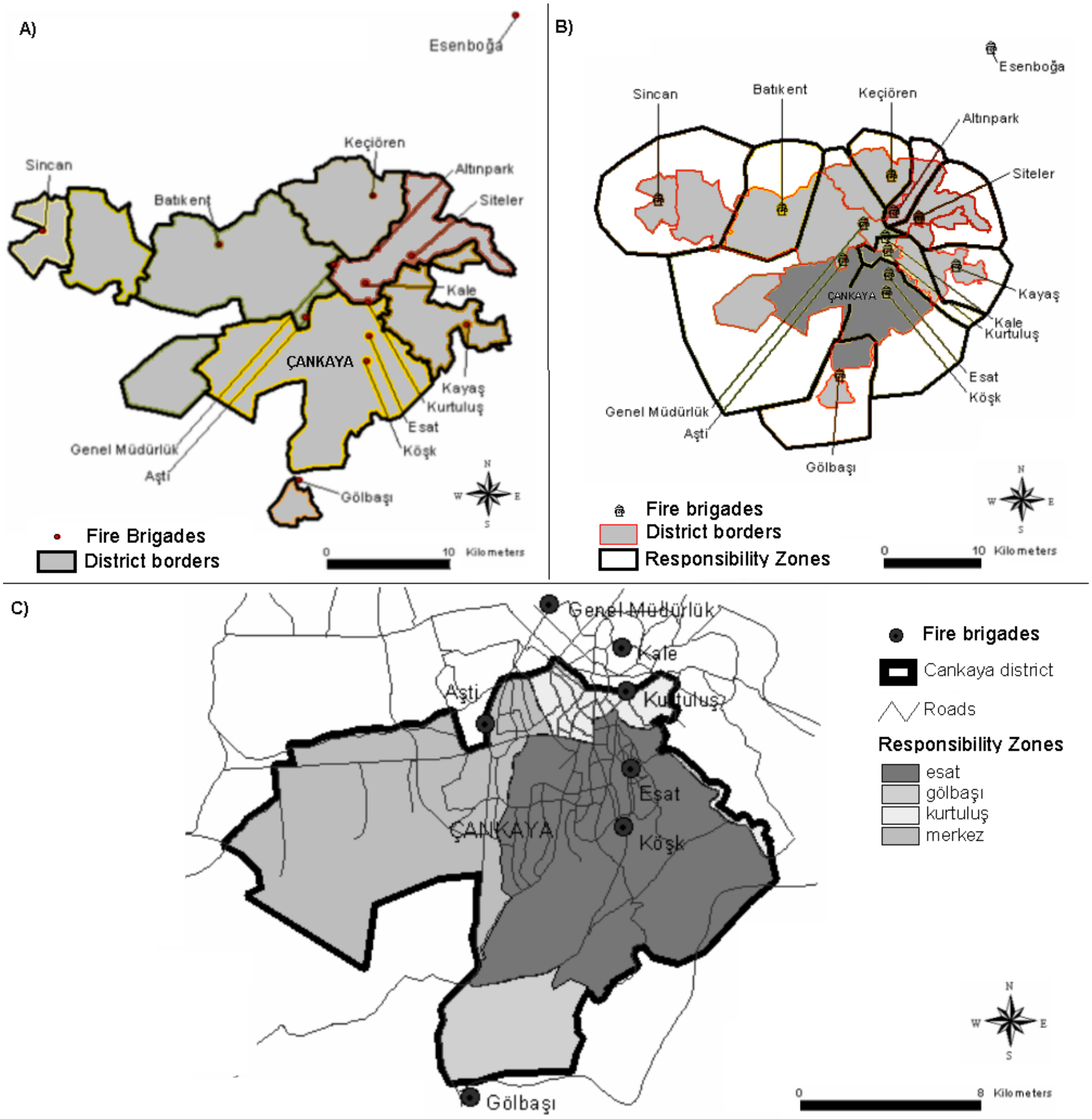

Figure 2: Fire brigades and responsibility zones in Ankara

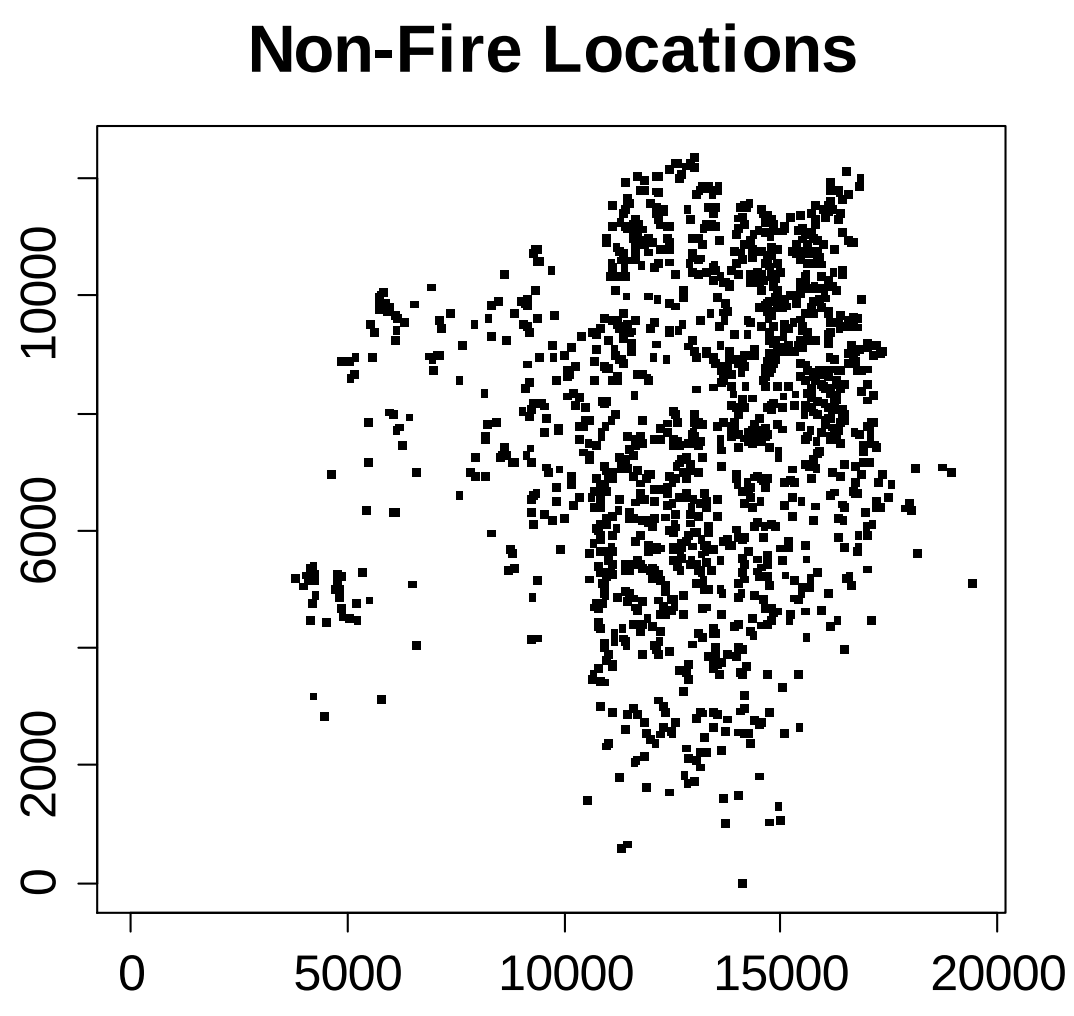

Figure 3: The plots of the locations of non-fire cases for all years combined.

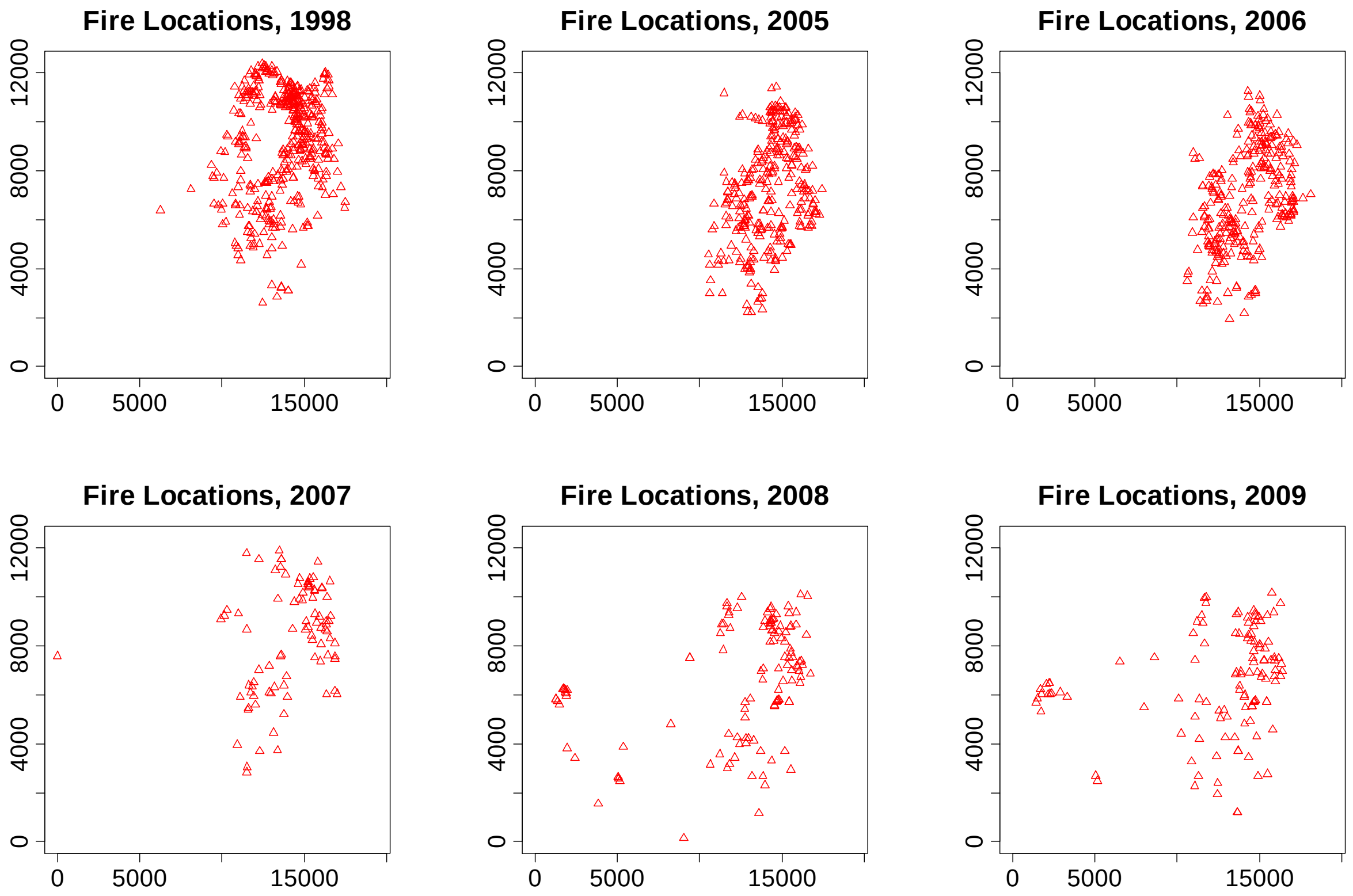

Figure 4: The scatter plots of the fire locations for each year for 1998, and 2005-09,

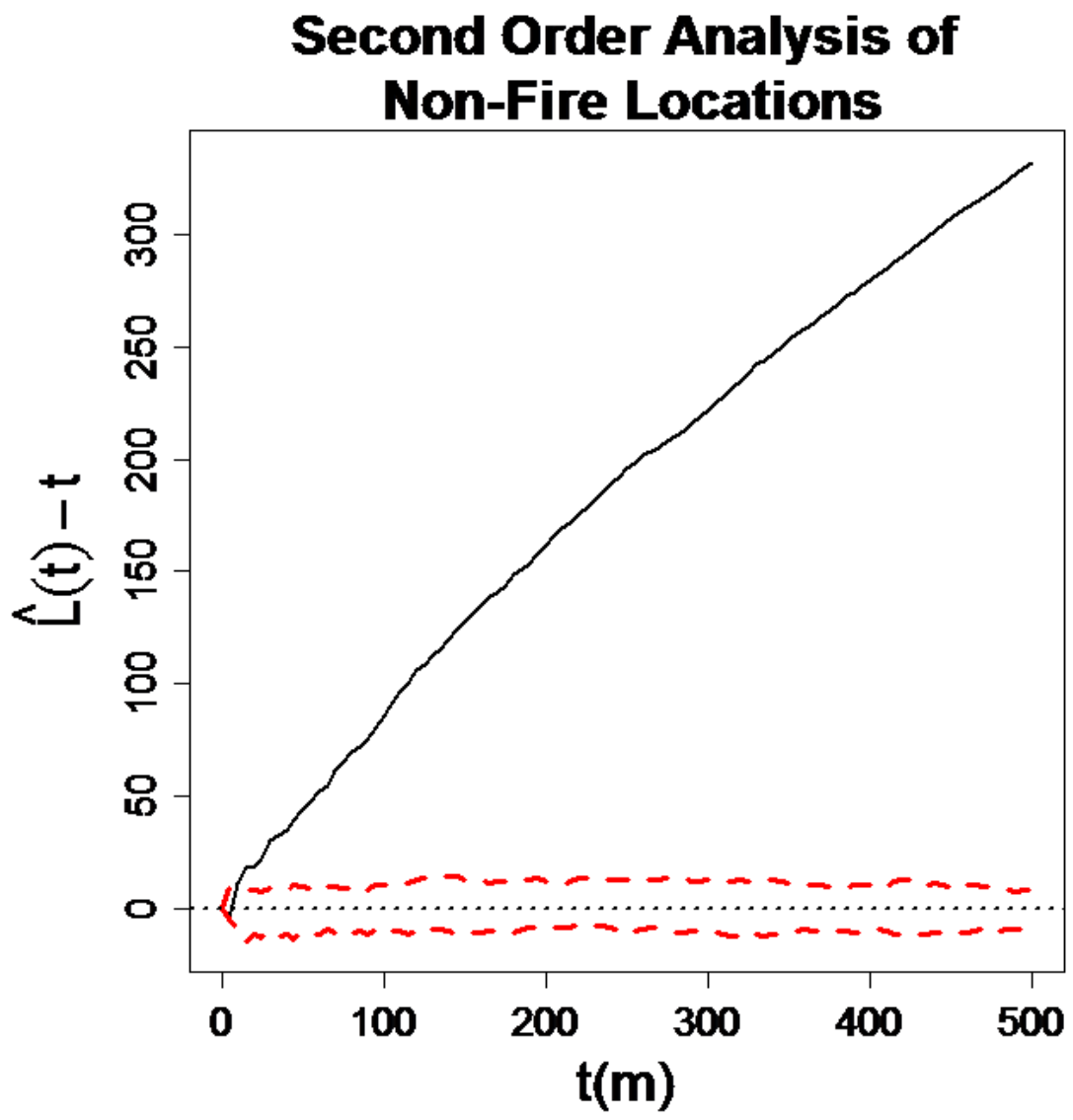

Figure 5: Second-order analysis of non-fire locations: Function plotted is Ripley's univariate $L$-function $\hat{L}(t)-t$ as a function of distance $t$ in 0-500 meters. The dotted horizontal line at 0 is the expected value under CSR, while the dashed lines around 0 are 99% confidence bands of non-fire residences under CSR.

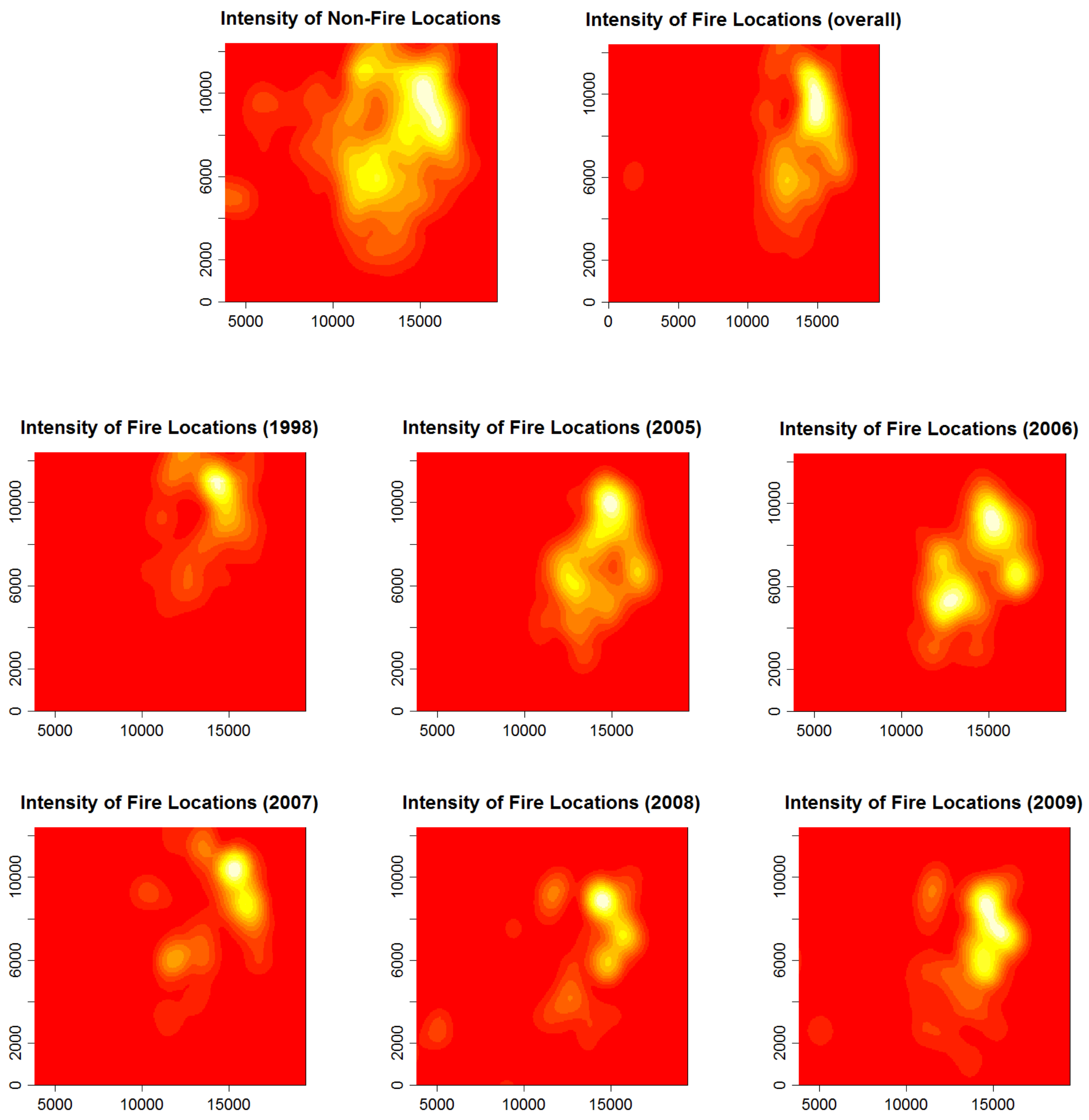

Figure 6: The intensity plots for non-fire cases (i.e., residences which did not experience fire) and for fire cases (with overall data and for each year).The intensity value increases from red to white

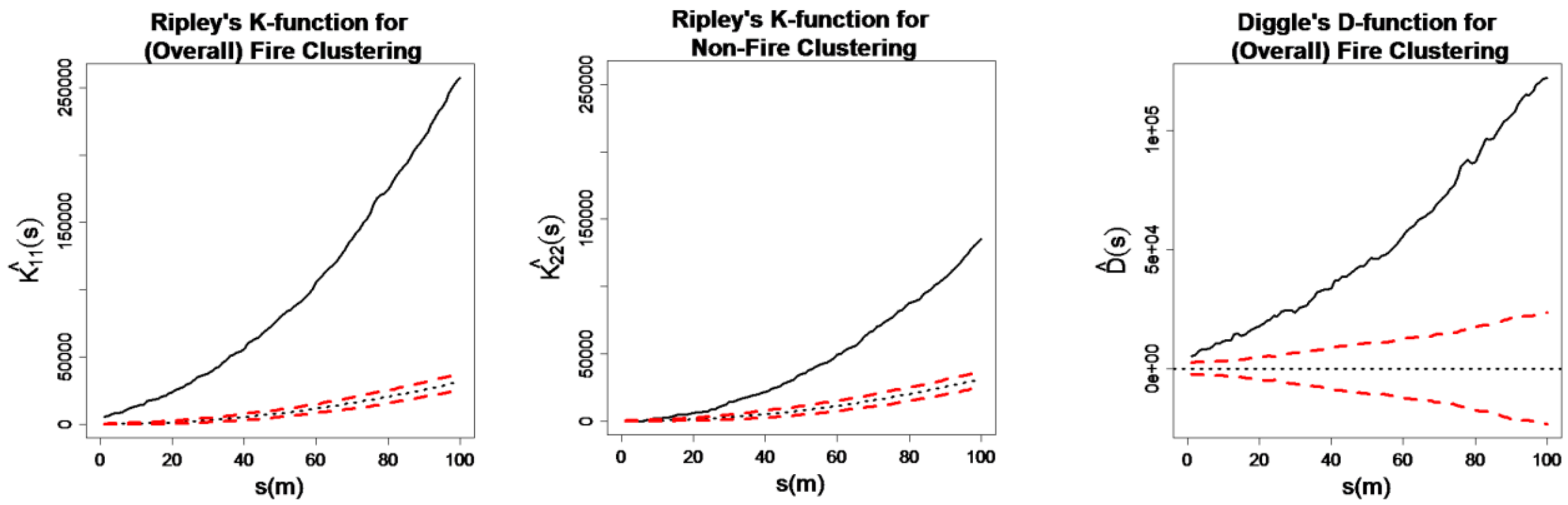

Figure 7: Second-order analysis of fire data. Functions plotted are Ripley's $K$-functions $\hat{K}_{ii}(s)$ with $i=1$ for fire and $i=2$ for non-fire cases (left and middle) and Diggle's $D$-function $\hat{D}(s)=\hat{K}_{11}(s)-\hat{K}_{22}(s)$ (right) for the overall data. The dashed lines in the left two plots are plus and minus two standard errors of $\hat{K}_{ii}(s)$ under RL of fire and non-fire cases and the dashed/dotted lines are for, $2\pi s^2$, the theoretical value of $\hat{K}_{ii}(s)$ and in the left plot are plus and minus two standard errors of $\hat{D}(s)$ under RL of fire and non-fire cases.

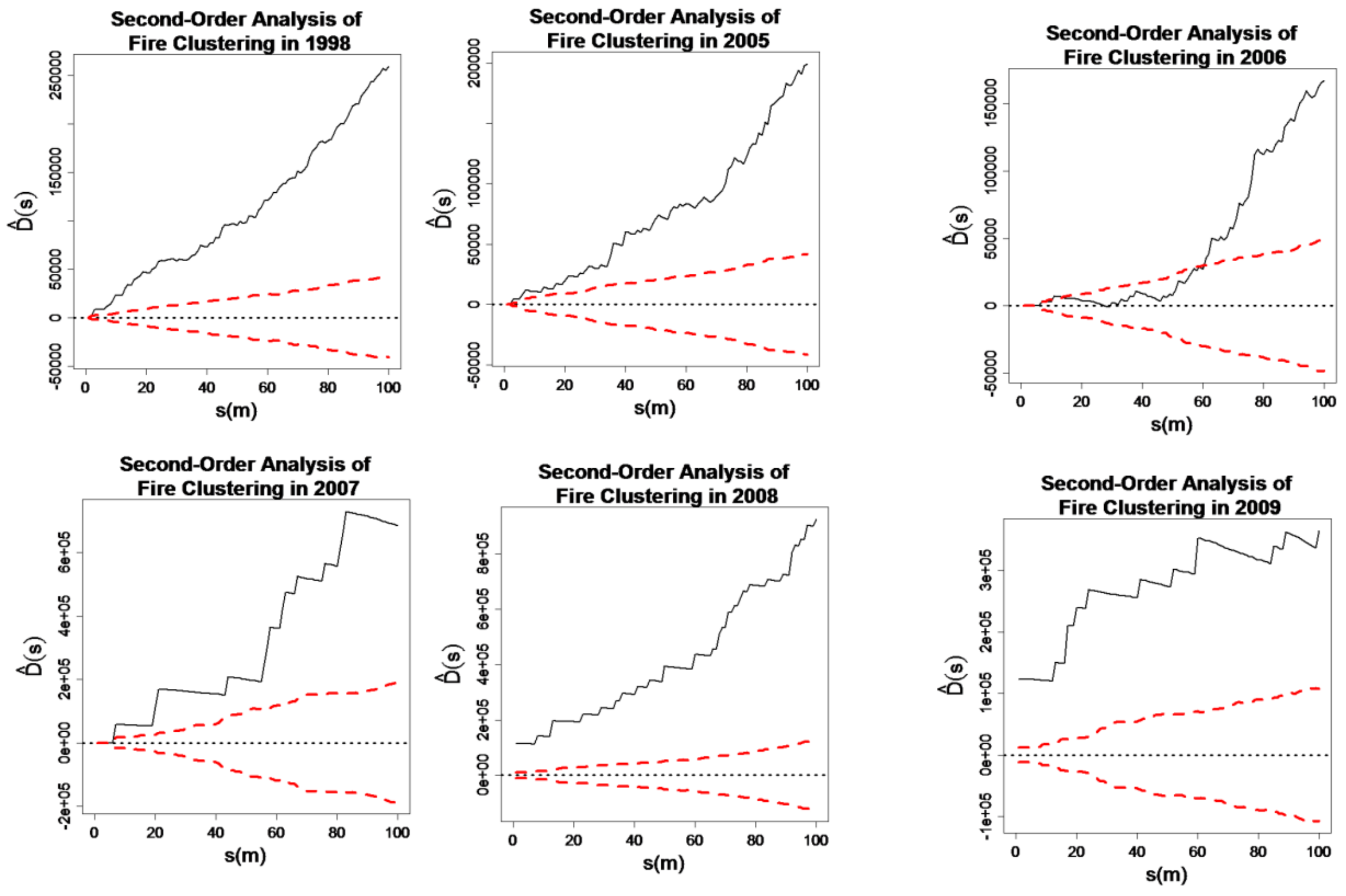

Figure 8: Second-order analysis of fire data. Functions plotted are Diggle's $D$-functions $\hat{D}(s) = \hat{K}_{11}(s) - \hat{K}_{22}(s)$ with $i = 1$ for fire and $i = 2$ for non-fire pattern for each year. The dashed lines around 0 are plus and minus two standard errors of $\hat{D}(s)$ under RL of fire and non-fire cases

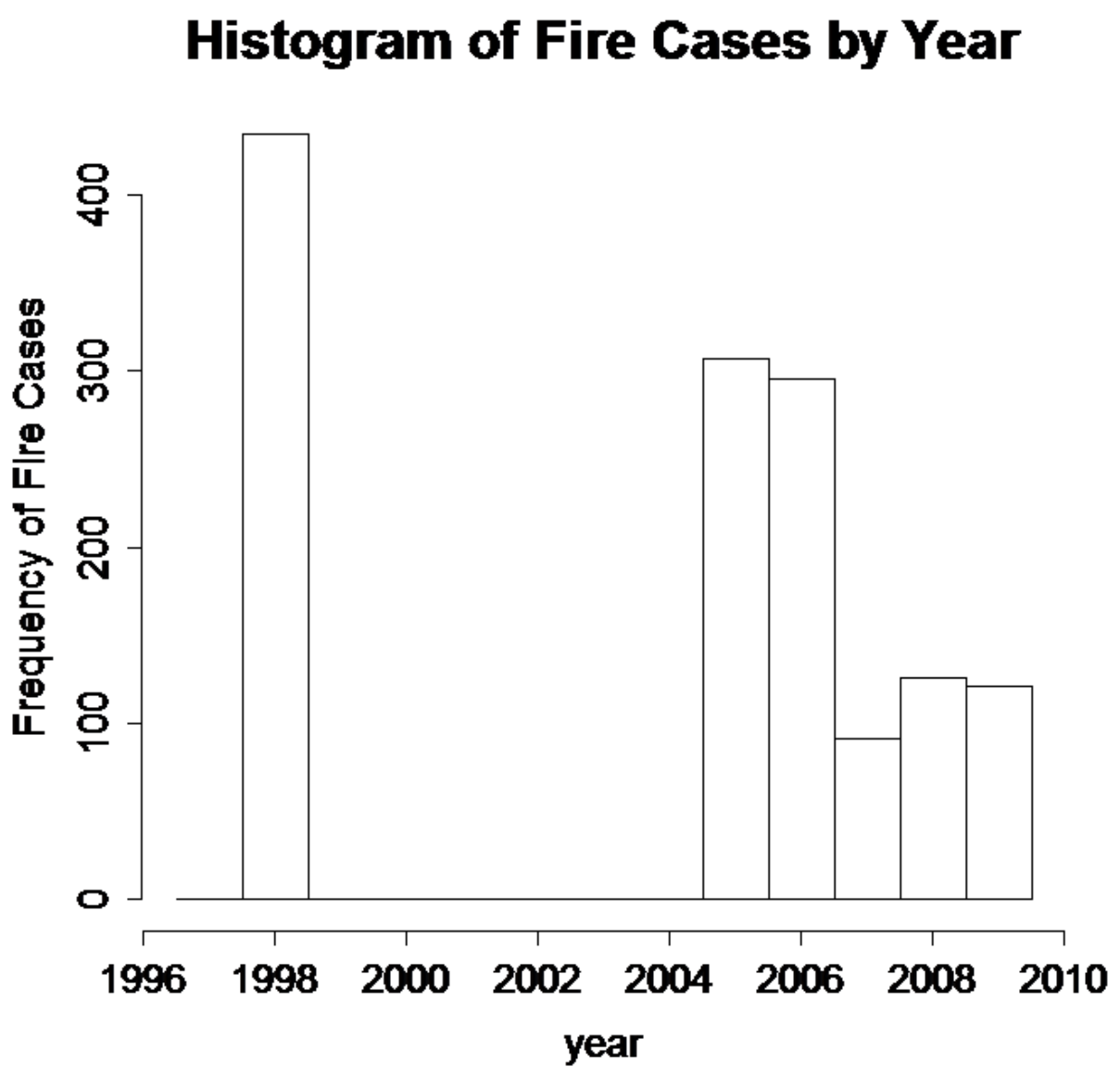

Figure 9: Frequency histogram of the fire cases by year (left) and by moth for the three years 2007-2009 combined (right)

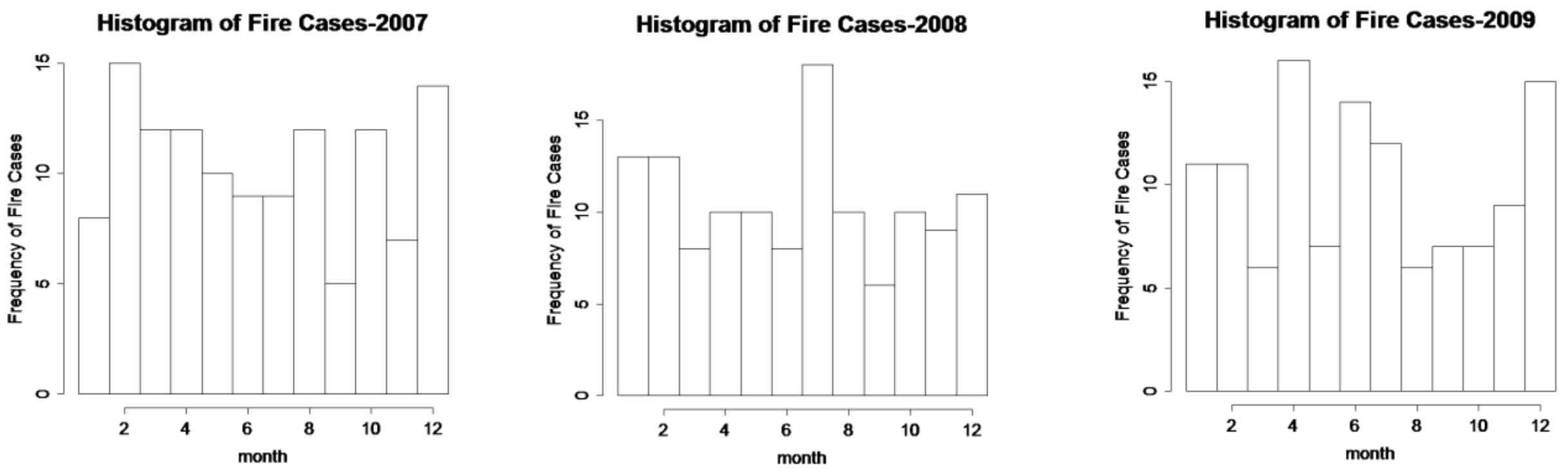

Figure 10: Frequency histogram of the fire cases by month for years 2007-2009

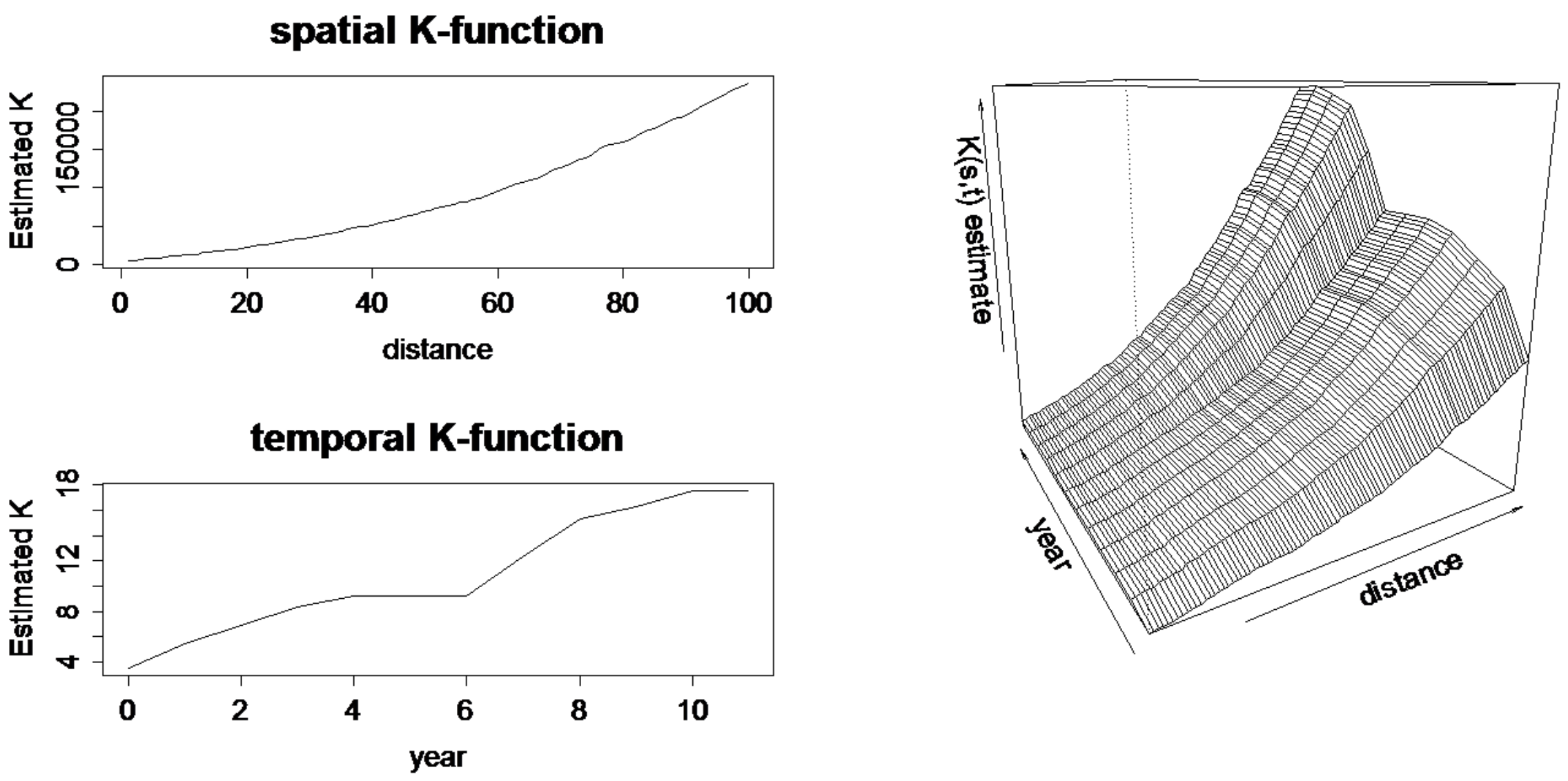

Figure 11: The spatial (left top) and temporal (left bottom) $K$-functions and the perspective plot of $K(s,t)$ estimate (right) as a function of year and distance

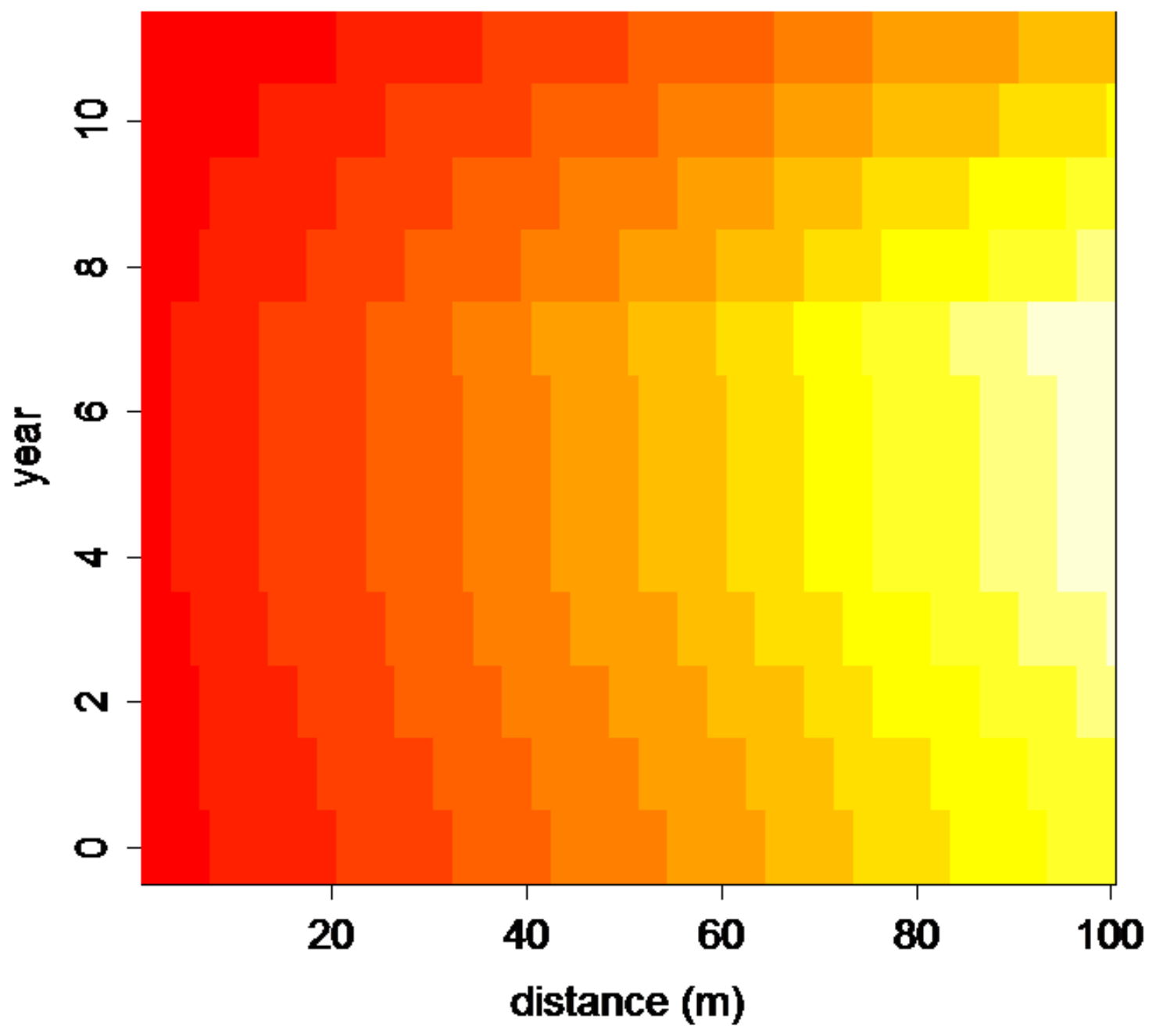

Figure 12: The color-coded grid plot of the standard error matrix for space-time clustering for the distance values from 1-100 m and year differences from 0-11 years. Entry $(i, j)$ of the matrix is the standard error at $s_i, t_j$ (See [27] for details). The standard error values increase from red to white

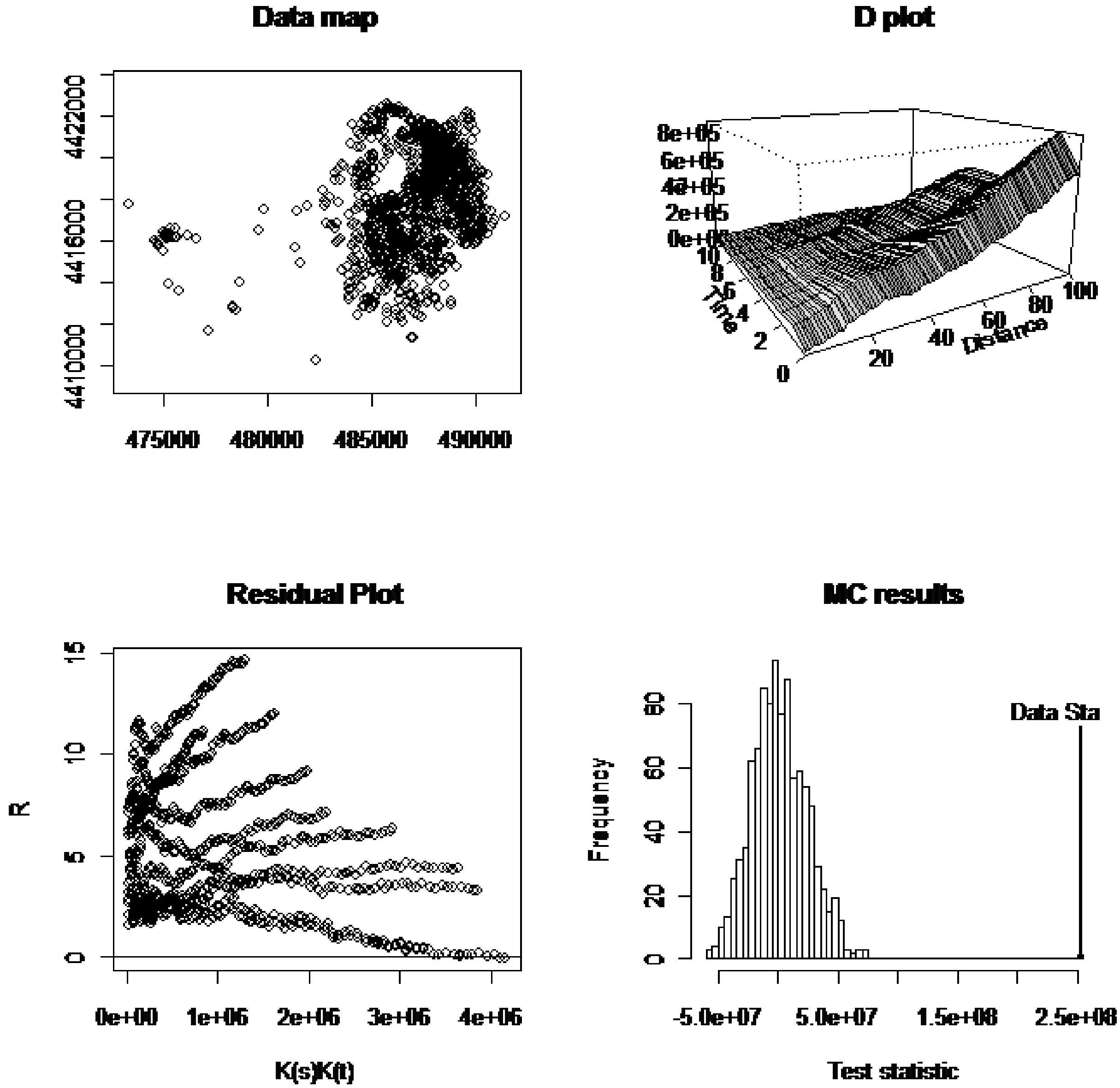

Figure 13: Diagnostic plots for space-time clustering. The four plots are the spatial map of the data points (top left), a perspective plot of the difference between spatio-temporal $K$-function and the product of the spatial and temporal $K$-functions (top right), the standardized residuals against the product of the spatial and temporal $K$-functions (bottom left) and histogram of the test statistics (bottom right), where the statistic for the data is indicated with a vertical line.

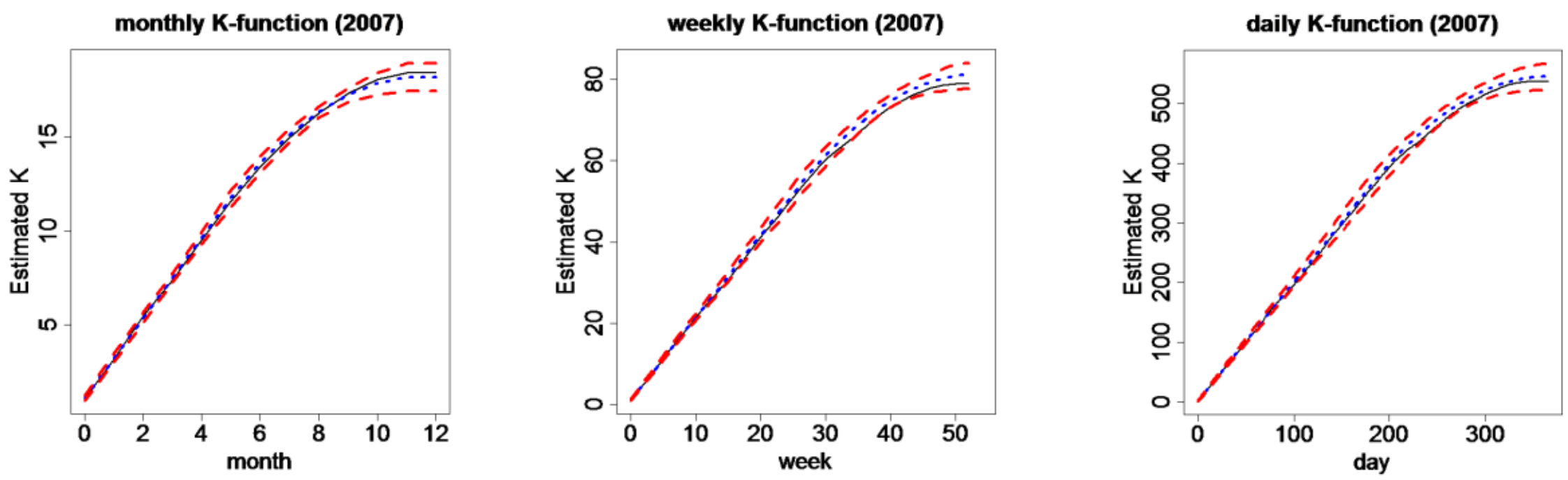

Figure 14: The temporal $K$-functions in year 2007 for time differences measured in month, week, and day. In each plot, solid line is the estimate based on data, the dotted lines are the estimated expected value based on uniform distribution of fires in time, and dashed lines are the 95% confidence limits

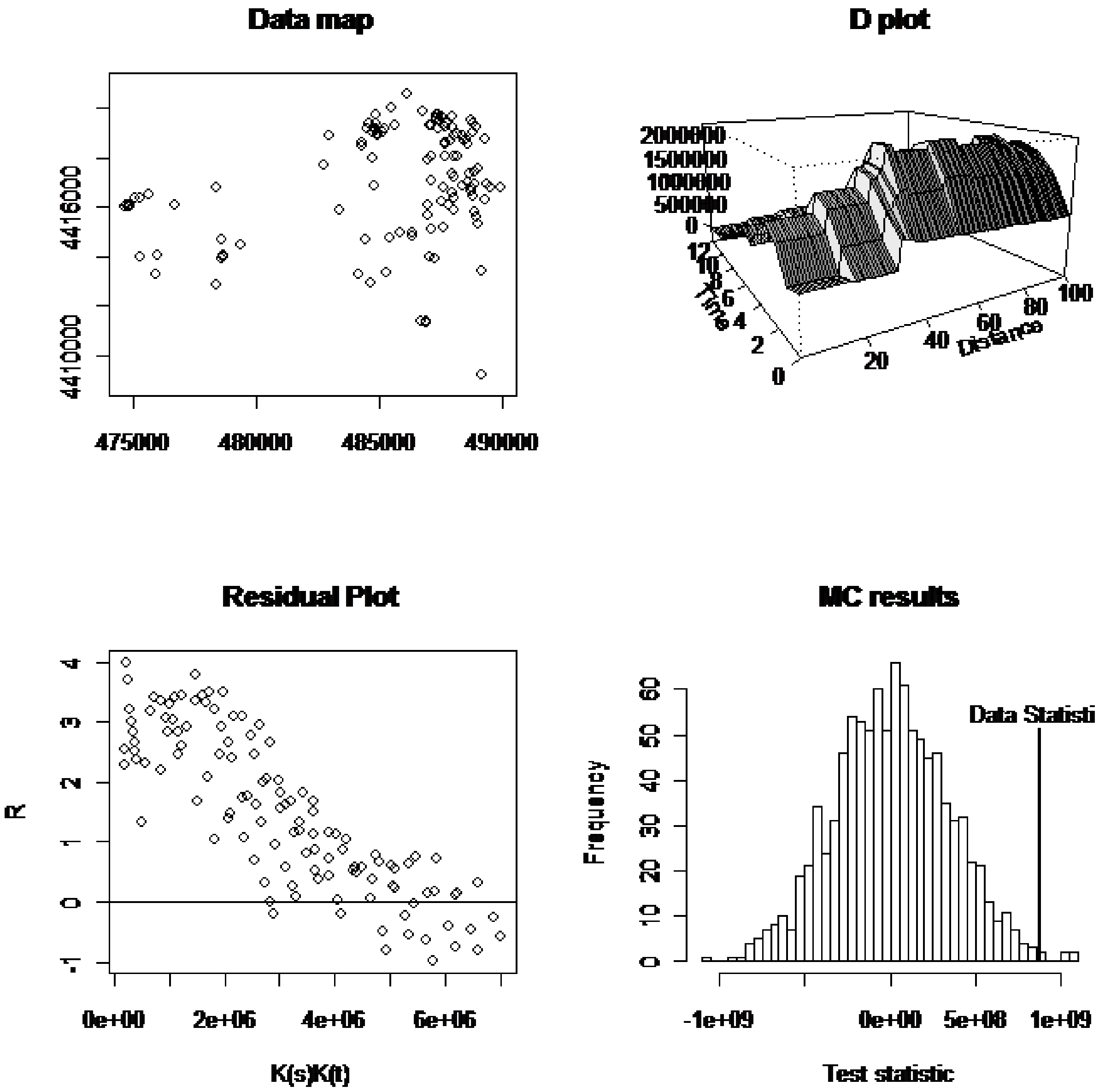

Figure 15: Diagnostic plots for space-time clustering for time measured in months in year 2007. The four plots are as in Figure 13

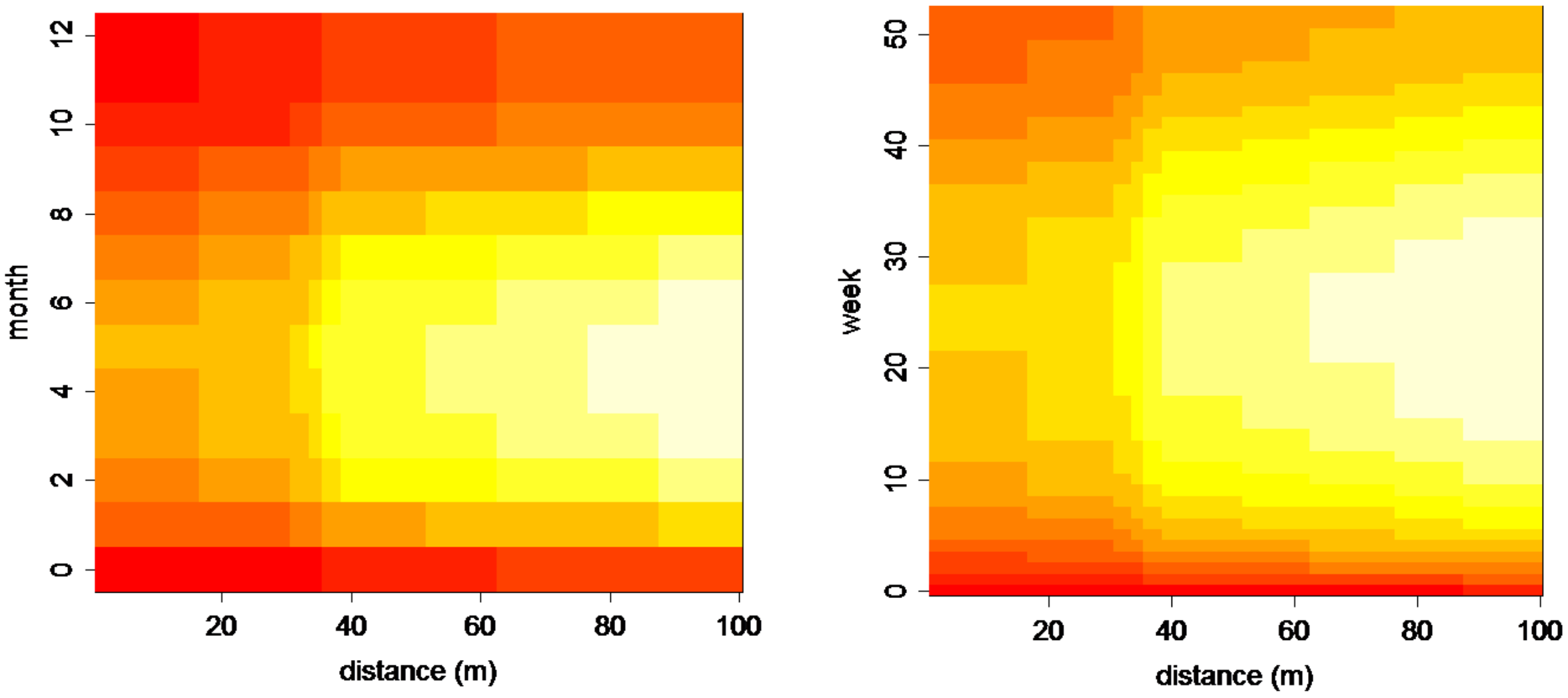

Figure 16: The color-coded grid plot of the standard error matrices for space-time clustering for the distance values from 1-100 m and months (left) and weeks (right) in 2007. The standard error values increase from red to white

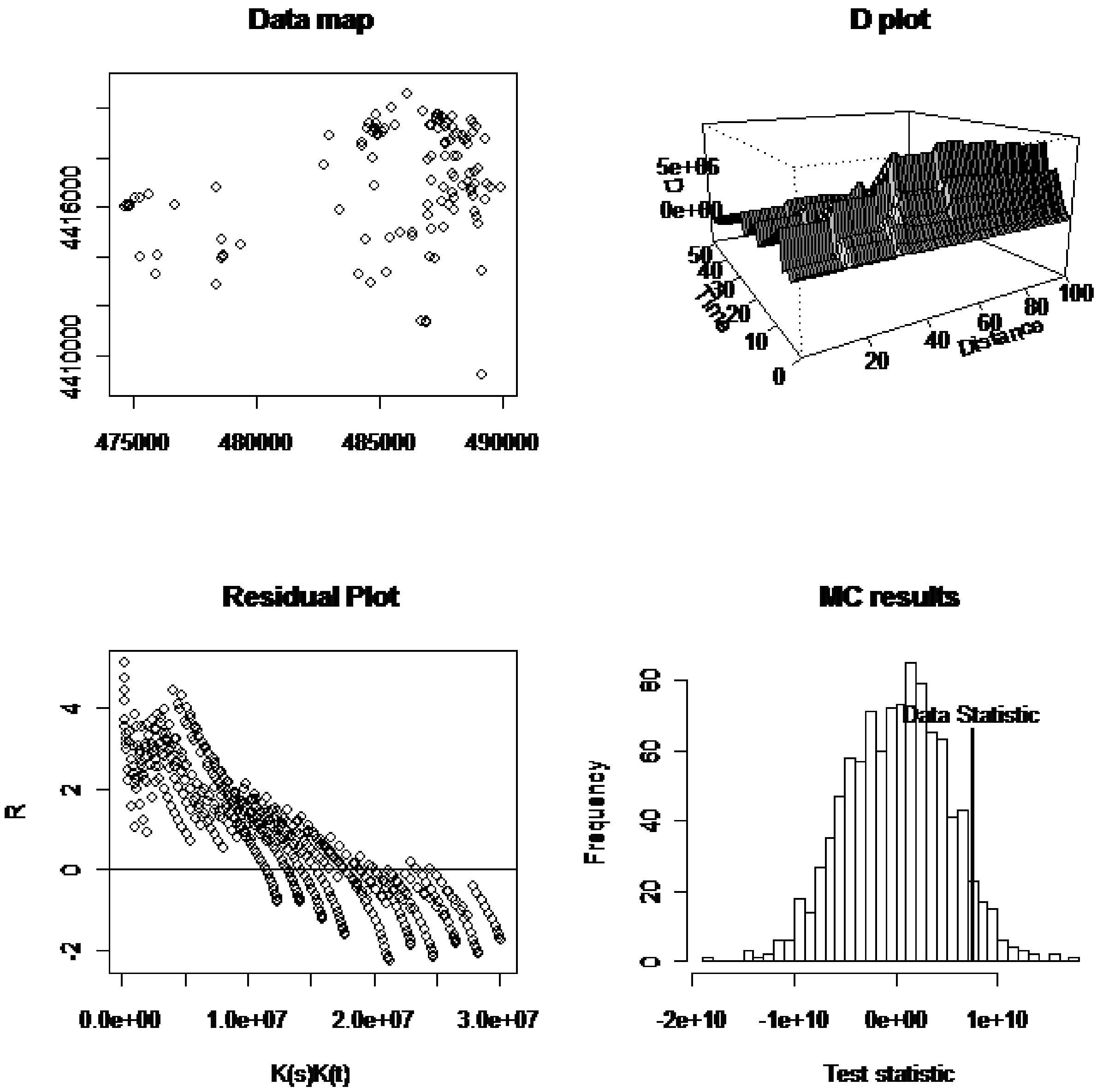

Figure 17: Diagnostic plots for space-time clustering for time measured in weeks in year 2007. The four plots are as in Figure 13

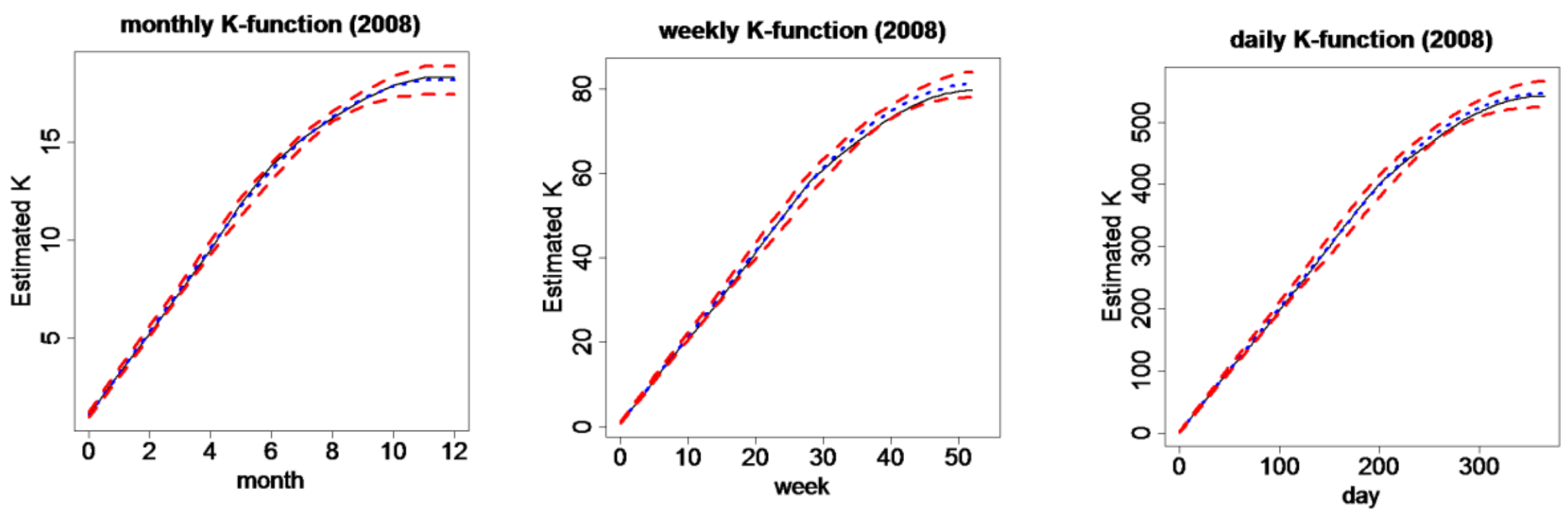

Figure 18: The temporal $K$-functions in year 2008 for time differences measured in month, week, and day. In each plot, solid line is the estimate based on data, the dotted lines are the estimated expected value based on uniform distribution of fires in time, and dashed lines are the 95% confidence limits.

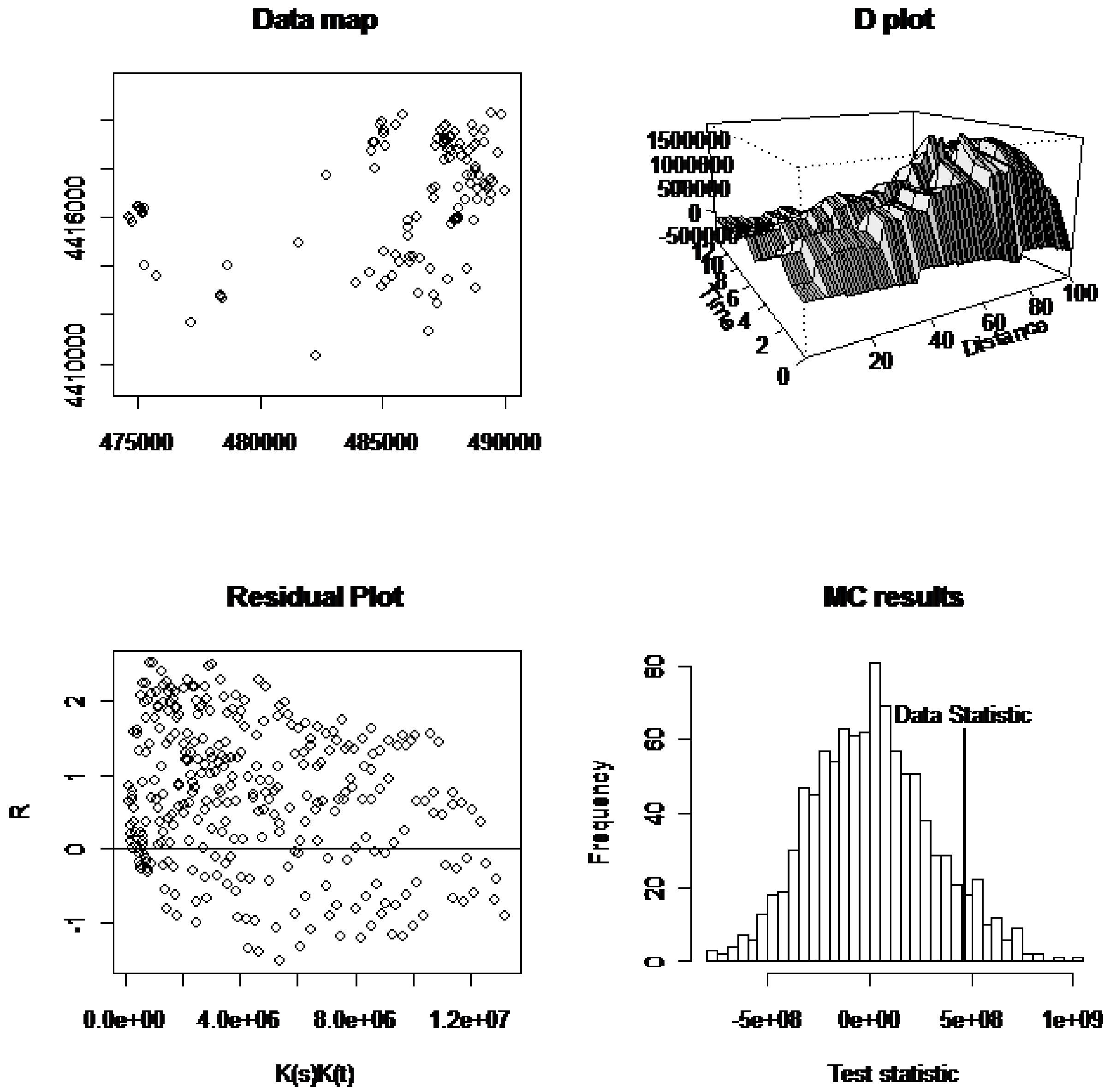

Figure 19: Diagnostic plots for space-time clustering for time measured in months in year 2008.The four plots are as in Figure 13

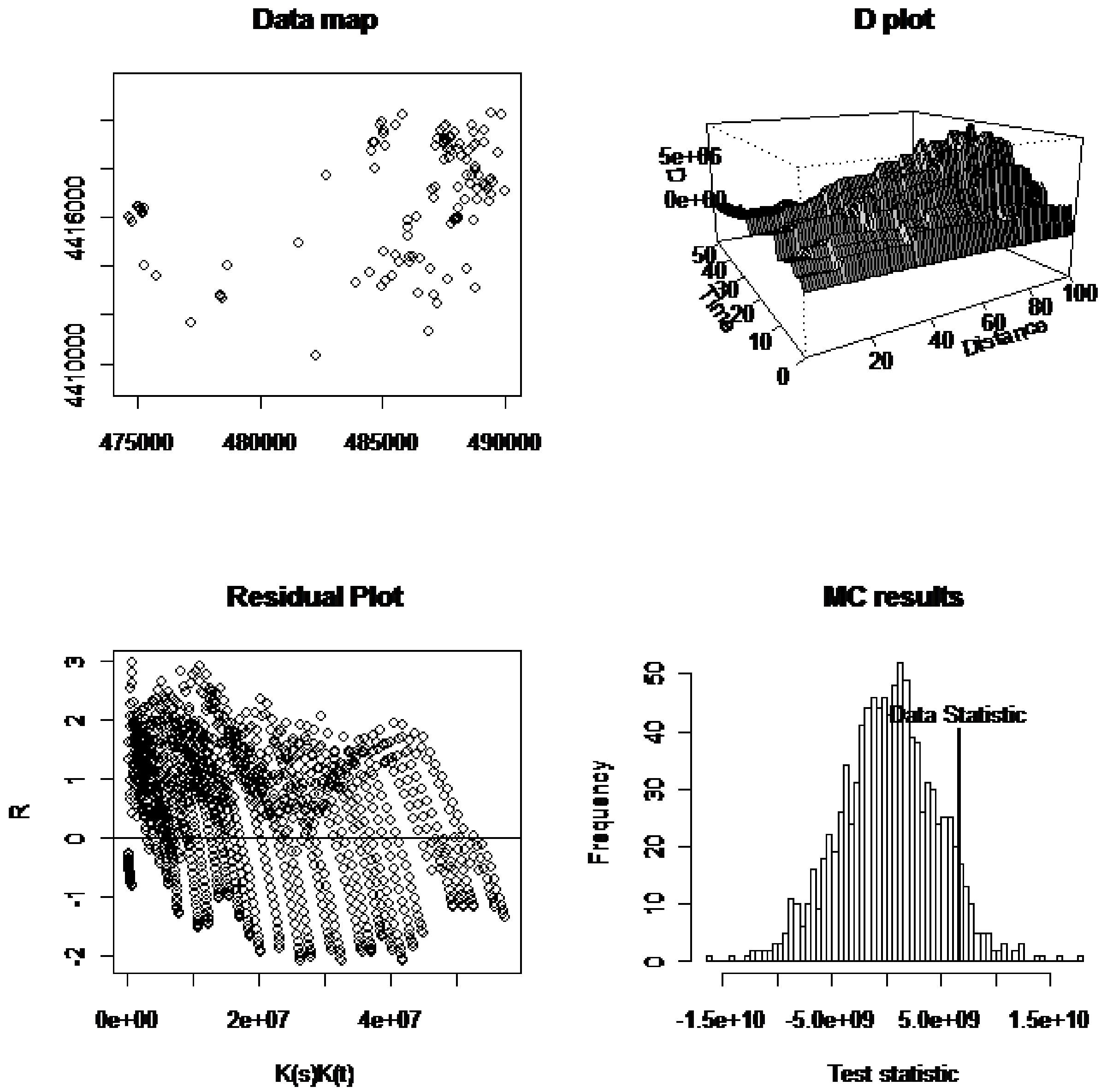

Figure 20: Diagnostic plots for space-time clustering for time measured in weeks in year 2008 (The four plots are as in Figure 13)

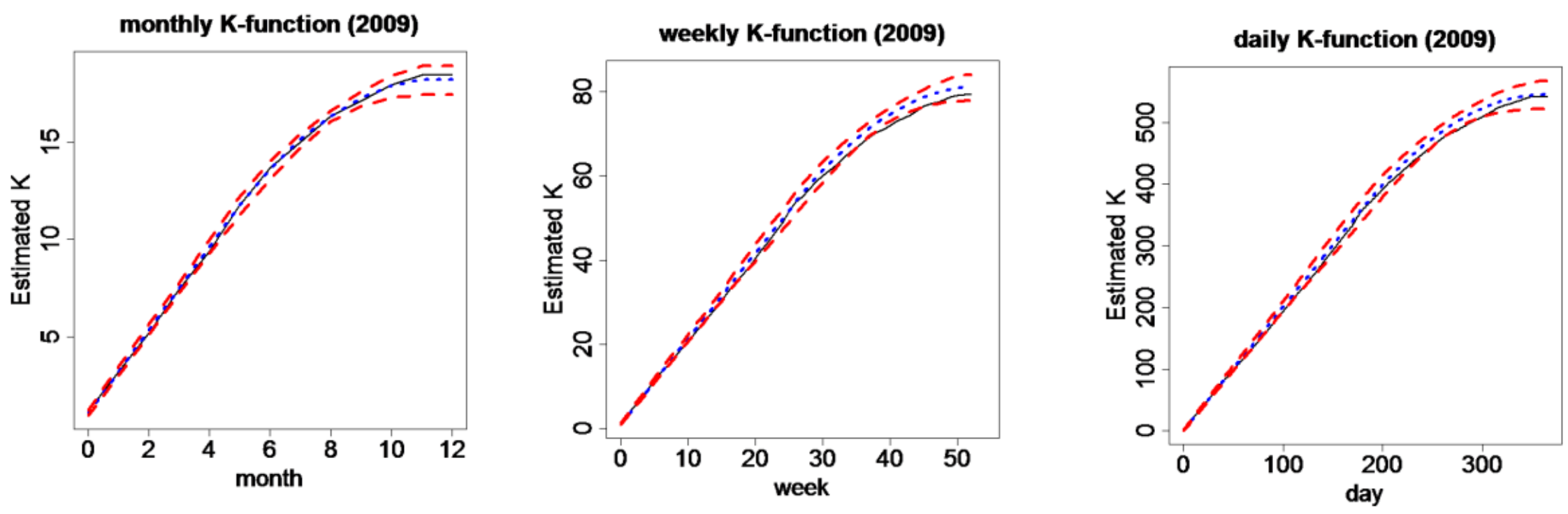

Figure 21: The temporal $K$-functions in year 2009 for time differences measured in month, week, and day. In each plot, solid line is the estimate based on data, the dotted lines are the estimated expected value based on uniform distribution of fires in time, and dashed lines are the 95% confidence limits

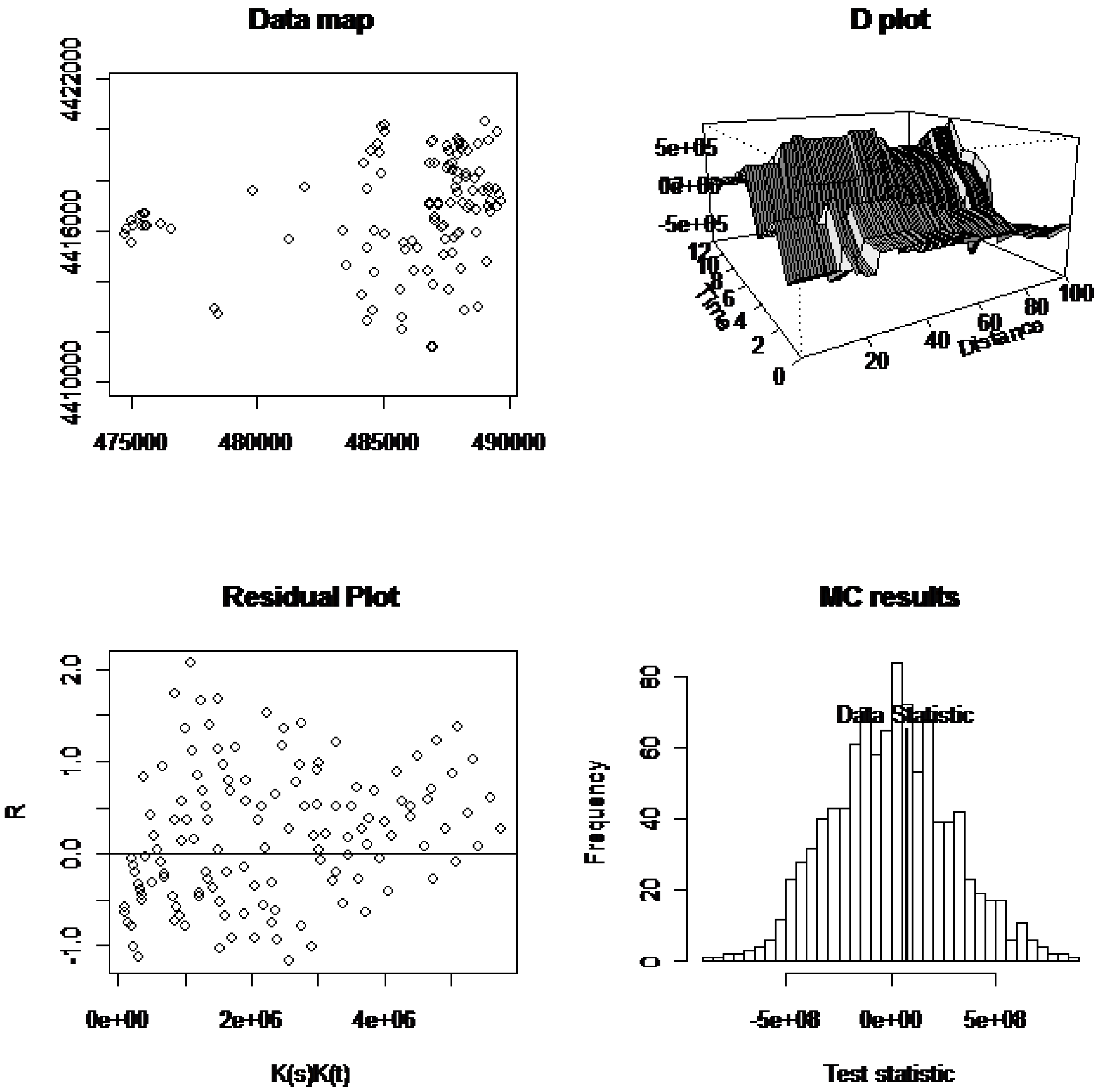

Figure 22: Diagnostic plots for space-time clustering for time differences measured in months in year 2009. The four plots are as in Figure 13

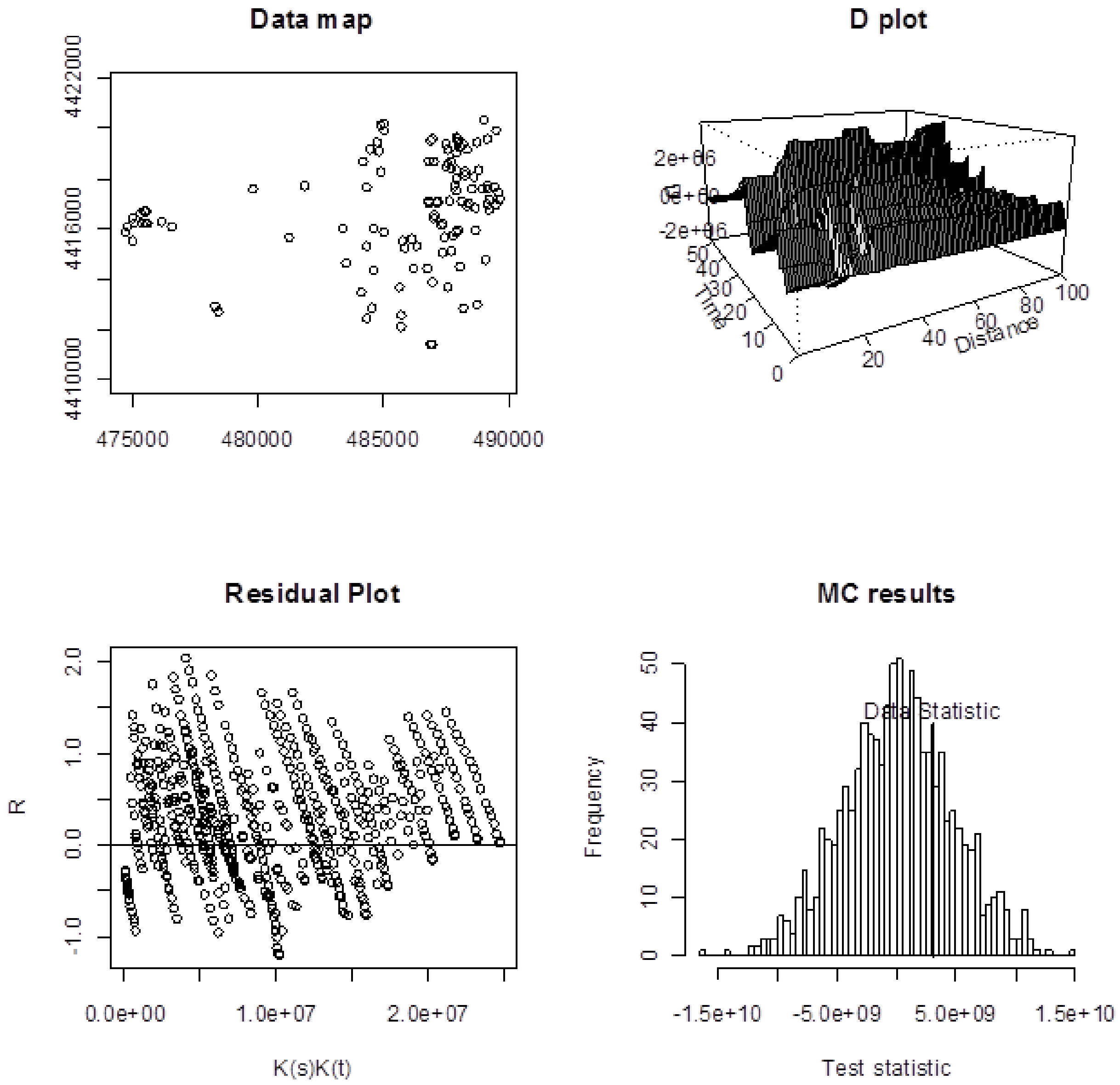

Figure 23: Diagnostic plots for space-time clustering for time differences measured in weeks in year 2009. The four plots are as in Figure 13